\documentclass[aps, pra, onecolumn, amsmath, amssymb, superscriptaddress, nofootinbib]{revtex4-1}

\makeatletter
\def\p@subsection{}
\def\p@subsubsection{}
\makeatother

\usepackage[utf8]{inputenc}
\usepackage{graphicx}
\usepackage{units}
\usepackage{color}
\usepackage[pdftex, colorlinks=true, linkcolor=myblue, citecolor=myblue,
urlcolor=myblue]{hyperref}
\usepackage{times}
\usepackage{comment}
\usepackage{graphicx}
\usepackage{amsmath, calc}
\usepackage{mathrsfs}
\usepackage{amsfonts}
\usepackage{amssymb}
\usepackage{bm}
\usepackage{bbm}
\usepackage{color}
\usepackage{array}
\usepackage{units}

\definecolor{myblue}{rgb}{0,0,1}
\definecolor{myred}{rgb}{1,0,0}

\newcommand{\bra}[1]{\langle #1|}
\newcommand{\ket}[1]{|#1\rangle}

\begin{document}


\title{Unconventional edge states in a two-leg ladder}


\author{C. A. Downing} 
\email{Corresponding author: c.a.downing@exeter.ac.uk}
\affiliation{Department of Physics and Astronomy, University of Exeter, Exeter EX4 4QL, United Kingdom}

\author{L. Martín-Moreno}
\affiliation{Instituto de Nanociencia y Materiales de Aragón (INMA), CSIC-Universidad de Zaragoza, 50009 Zaragoza, Spain}
\affiliation{Departamento de Física de la Materia Condensada, Universidad de Zaragoza, 50009 Zaragoza, Spain}

\author{O. I. R. Fox}
\affiliation{Department of Physics and Astronomy, University of Exeter, Exeter EX4 4QL, United Kingdom}




\begin{abstract}
\noindent \\
\\
\textbf{Abstract}\\
Some popular mechanisms for restricting the diffusion of waves include introducing disorder (to provoke Anderson localization) and engineering topologically non-trivial phases (to allow for topological edge states to form). However, other methods for inducing somewhat localized states in elementary lattice models have been historically much less studied. Here we show how edge states can emerge within a simple two-leg ladder of coupled harmonic oscillators, where it is important to include interactions beyond those at the nearest neighbour range. Remarkably, depending upon the interplay between the coupling strength along the rungs of the ladder and the next-nearest neighbour coupling strength along one side of the ladder, edge states can indeed appear at particular energies. In a wonderful manifestation of a type of bulk-edge correspondence, these edge state energies correspond to the quantum number for which additional stationary points appear in the continuum bandstructure of the equivalent problem studied with periodic boundary conditions. Our theoretical results are relevant to a swathe of classical or quantum lattice model simulators, such that the proposed edge states may be useful for applications including waveguiding in metamaterials and quantum transport.
\end{abstract}


\maketitle



\section{Introduction}
\label{Sec:Introduction}

The theory of the localization of electrons due to disorder, as created by randomness in the atomic lattice, is a towering achievement in condensed matter physics~\cite{Anderson1958, Thouless1974}. This phenomena of an absence of diffusion can be observed throughout waves physics, including in acoustics, electromagnetics and photonics~\cite{Wiersma2013, Segev2013}. Localization in perfectly ordered systems is also possible, most famously thanks to Wannier-Stark localization whereby an applied electric field breaks the periodicity of the lattice~\cite{Wannier1962, Emin1987}. Latterly, the rise of topological matter has celebrated the presence of highly localized states hosted by certain lattices, including topologically-protected edge states which exist in the gaps formed between bands filled with extended states~\cite{Arikawa2009, Hasan2010, Asboth2016}. Furthermore and perhaps counterintuitively, the reality of so-called bound states in the continuum -- in the broadest sense, localized states existing within a continuous spectrum of delocalized waves -- have also been demonstrated in a series of photonic architectures~\cite{Hsu2016, Azzam2020, Sadreev2021, Koshelev2023} nearly a century after the original idea of von Neumann and Wigner~\cite{Neumann1929}.

Within condensed matter theory, the consideration of couplings going beyond the nearest-neighbour approximation is a perennially popular activity within standard treatments of lattice models~\cite{Fairbairn1968, Fisher1981, Landau1983, Han1994, Chandler2016}. Laterly, spiritually similar theoretical studies have been performed for acoustical, optical, and phononic platforms amongst others, with interesting implications for topological phases and for the creation of rather exotic dispersion relations~\cite{Laha2017, Mittal2018, Chen2021, Saroka2021, Aristov2022, Wang2022, Kazemi2023}. Experimentally, physical systems with non-negligible interactions beyond those of the nearest-neighbour type have been reliably produced, in some cases with tunability, using atomic spins inside a cavity~\cite{Periwal2021};  acoustic metamaterials~\cite{ZhuGao2022}; elastic metamaterials~\cite{Martinez2021, Bossart2023, Rajabpoor2023}; coupled optical cavities~\cite{Caselli2015, Senanian2023}, and exciton-polariton condensates in microcavities~\cite{Alyatkin2020, Dovzhenko2023}. However, thus far the creation of edge states due to the inclusion of longer-range coupling has been mostly ignored, perhaps because of the seemingly contradictory concepts at play.

\begin{figure*}[tb]
 \includegraphics[width=\linewidth]{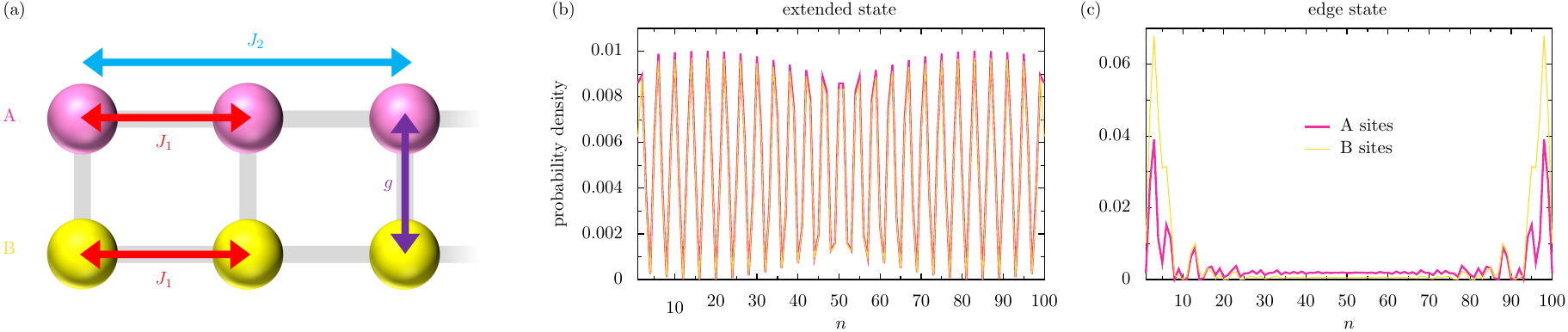}
 \caption{ \textbf{Unconventional edge states in the bosonic two-leg ladder model.} Panel (a): a sketch of the two-leg ladder system under consideration, composed of two linear chains (denoted $A$ and $B$) of quantum harmonic oscillators. Each oscillator is coupled (with the strength $g$, purple arrows) to its partner across the rung. The upper chain ($A$, pink balls) exhibits both nearest-neighbour coupling (with the strength $J_1$, red arrows) and next-nearest-neighbour coupling (with the strength $J_2$, cyan arrows). The lower chain ($B$, yellow balls) displays nearest-neighbour coupling (with $J_1$, red arrows) only. The rungs are equally spaced at intervals of the distance $d$ along the ladder. Panel (b): the probability density of a typical extended state over the $2N$ sites of the two-leg ladder. The results for the $A$ and $B$ sites $n$ are shown with pink and yellow lines respectively. Panel (c): the equivalent profile of a typical edge state. In the figure, we consider a two-leg ladder with a total of $2N = 200$ oscillators, and the couplings $g = 3 J_1/2$ and $J_2 = J_1$. The scaled participation ratio $\mathrm{PR} \left( m \right) / 2 N$ of the states considered is the rather typical $\simeq 0.675$ in panel (b) and perhaps surprisingly just $\simeq 0.171$ in panel (c).}
 \label{wfs}
\end{figure*}

Here we propose exploiting longer-range couplings in order to generate edge states -- without recourse to disorder, aperiodicity or topological non-trivialities. As a prototypical system, we employ a bosonic two-leg ladder model~\cite{Troyer1996, Donohue2001, Orignac2001, Robinson2012, Wei2014, Ye2022}, which has already been explored experimentally to great effect using ultracold atoms for example~\cite{Atala2014, Greschner2015, Tai2017, An2017, Zhu2022, Xiao2023}. While previous studies of so-called `biased' two-leg ladders have considered either interchain energy detunings across the two legs of the ladder or different two-body onsite interaction terms~\cite{Deng2015, Qiao2021, Padhan2023, Fan2023}, here we consider a bias in the range of the couplings themselves. We suppose that the two-leg ladder is composed of an upper chain A [marked in pink in Fig.~\ref{wfs}~(a)] which exhibits both nearest-nearest and next-nearest-neighbour couplings, and a lower chain B [depicted in yellow in panel~(a)] which only sustains nearest-neighbour interactions. Perhaps surprisingly, this coupling imbalance raises the possibility for edge states to emerge amongst the expected plethora of extended states which are spread throughout the ordered lattice, as implied in the probability density plots already previewed in Fig.~\ref{wfs}~(b, c). Notably, the next-nearest-neighbour coupling introduces additional stationary points in the bandstructure (found in the continuum limit, after employing periodic boundary conditions) which occur for a certain quantum number $q$, which is associated with a definite eigenfrequency~\cite{Brillouin1953}. This particular bandstructure eigenfrequency is in turn exactly the energy of the edge state (found after diagonalization of the analogous finite-sized lattice problem with the implied open boundary conditions), which suggests a sort of bulk-edge correspondence between the generation of auxiliary stationary points and the appearance of edge states. Furthermore, the discovered edge states appear inside the notional band of extended states, such they may be thought of as a kind of bound state in the continuum, a flavour of state which is of a high current interest within modern condensed matter physics~\cite{Hsu2016, Azzam2020, Sadreev2021, Koshelev2023}.

The Hamiltonian operator $\hat{H}$ of the bosonic two-leg ladder sketched in Fig.~\ref{wfs}~(a) is composed of just three parts
\begin{equation}
\label{eq:Haxcsdfdsfdsfsdfvcxvmy}
 \hat{H} =  \hat{H}_A +  \hat{H}_B +  \hat{H}_{A-B},
\end{equation}
where $\hat{H}_A$ describes the $N$ coupled harmonic oscillators comprising the chain A subsystem [depicted as pink spheres in Fig.~\ref{wfs}~(a)] via the following Hamiltonian (we have set $\hbar = 1$ here and throughout)
\begin{equation}
\label{eq:sdfsdfsfd}
\hat{H}_A = \sum_{n=1}^N \omega_0 a_n^\dagger a_n + \sum_{n=1}^{N-1} J_1 \left( a_{n+1}^\dagger a_n + a_{n}^\dagger a_{n+1} \right)
 + \sum_{n=1}^{N-2} J_2 \left( a_{n+2}^\dagger a_n + a_{n}^\dagger a_{n+2} \right).
\end{equation}
The nearest-neighbour coupling strength $J_1 > 0$ [red arrow in Fig.~\ref{wfs}~(a)], and the next-nearest-neighbour coupling strength $J_2$ [cyan arrow] satisfies the inequality $0 \le J_2  \le J_1$. The $N$ oscillators in the chain B subsystem [yellow spheres in Fig.~\ref{wfs}~(a)], on the opposite side of the rung to their chain A counterparts, are governed by a standard nearest-neighbour tight-binding Hamiltonian $\hat{H}_B$, where
\begin{equation}
\label{eq:sdfsdssfdsfsfd}
\hat{H}_B = \sum_{n=1}^N \omega_0 b_n^\dagger b_n + \sum_{n=1}^{N-1} J_1 \left( b_{n+1}^\dagger b_n + b_{n}^\dagger b_{n+1} \right).
\end{equation}
The creation and annihilation operators $c_n^\dagger$ and $c_n$ create and destroy a bosonic excitation on the $n$-th rung of the ladder, where the two possible flavours of operator $c_n = \{ a_n, b_n \}$ refer to the site at the top or bottom of the $n$-th rung, and where the bosonic commutation relation $[ c_n, c_n^\dagger ] = 1$ is always observed. The common resonance frequency of all $2N$ oscillators throughout the two-leg ladder is $\omega_0$, although all of our results can easily be generalized to the case when the upper chain and lower chain have different resonance frequencies (this situation is briefly discussed later on in an appendix). Each oscillator is coupled to its adjacent partner along the rung (of the opposite flavour, $A$ or $B$) via the coupling Hamiltonian
\begin{equation}
\label{eq:sdfsdsssdfsdfsdffdsfsfd}
\hat{H}_{A-B} = \sum_{n=1}^{N} g \left( a_{n}^\dagger b_n + b_{n}^\dagger a_n \right).
\end{equation}
The inter-chain coupling strength $g > 0$ [purple arrow in Fig.~\ref{wfs}~(a)], such that the two dimensionless parameters of primary interest for the full model of Eq.~\eqref{eq:Haxcsdfdsfdsfsdfvcxvmy} are the next-nearest neighbour coupling ratio $J_2/J_1$, with a size between $0 \le J_2/J_1 \le 1$, and the inter-chain coupling ratio $g/J_1 \ge 0$. Together these coupling ratios control two important aspects of the overall model. Increasing $J_2/J_1$ allows for more stationary points in the continuum Hamiltonian bandstructure to emerge~\cite{Brillouin1953}, which raises the possibility for interference effects. For example, the group velocity and phase velocity may be in the same direction for one mode and in the opposite direction for another mode somewhere elsewhere in the Brillouin zone~\cite{Chen2021}. Meanwhile, increasing $g/J_1$ allows for more mixing between the chain A and chain B subsystems which, due to their different boundary conditions coming from their different coupling ranges, raises the possibility for novel states to be forged in the combined ladder system.

In what follows, we discuss the band theory of the total Hamiltonian $\hat{H}$ of Eq.~\eqref{eq:Haxcsdfdsfdsfsdfvcxvmy} in the infinitely long limit (and with periodic boundary conditions) in Sec.~\ref{Sec:Bandtheory}, which hints at novel states near to additional stationary points which arise due to interactions beyond nearest-neighbour when the coupling strength $J_1/4 \le J_2 \le J_1$. We consider the proper finite system defined by Eq.~\eqref{eq:Haxcsdfdsfdsfsdfvcxvmy} in Sec.~\ref{Sec:Finitesystems} (with its implied open boundary conditions), where we quantify the degree of localization of the eigenstates using the participation ratio. We also judge the character of the eigenvalue statistics using the mean value of the consecutive energy level spacings, which allows us to paint the extended state-to-edge state phase diagram of the model in Fig.~\ref{numer}. Some conclusions from our theoretical study are drawn in Sec.~\ref{Sec:Discussion}. We relegate to the appendices a complete treatment of the chain A subsystem model in the App.~\ref{Sec:chainA}, including a derivation of the analytic form of the density of states beyond nearest neighbour interactions, and an analysis of the chain B subsystem model in App.~\ref{Sec:chainB}, including a derivation of the analytic probability density function associated with the adjacent energy level spacings. Finally, some supplementary results for the full two-leg ladder are housed within App.~\ref{Sec:chainAB}, while App.~\ref{Sec:detune} briefly considers a small extension of that model to allow for interchain energy detunings.
\\


\section{Band theory}
\label{Sec:Bandtheory}

Let us start by considering the full two-leg ladder model in the limit of large system size ($N \to \infty$) and after employing periodic boundary conditions, which allows for an analytic treatment of the continuum theory. Our hope is for the continuum model to have some descriptive power over the finite-sized system, in a similar manner to how the so-called bulk-edge correspondence in the physics of topological matter suggests that a topological invariant can predict the presence or absence of topological edge states~\cite{Hasan2010, Asboth2016}. We introduce the pair of exponential Fourier transforms for the bosonic operators $a_n$ and $b_n$ appearing in the ladder model of Eq.~\eqref{eq:Haxcsdfdsfdsfsdfvcxvmy} as follows 
\begin{equation}
\label{eq:sdfssdfsdfdsssdfsdfsdffdsfsfd}
a_n = \frac{1}{\sqrt{N}} \sum_{q} \mathrm{e}^{\mathrm{i} n q d} a_q,
\quad\quad\quad\quad\quad
b_n = \frac{1}{\sqrt{N}} \sum_{q} \mathrm{e}^{\mathrm{i} n q d} b_q,
\end{equation}
with the quantum number $q = 2 \pi m / N d$, where the integer $m \in [ -N/2, +N/2 ]$ implies that the first Brillouin zone exists for $-\pi/d < q \le \pi / d$. The length scale $d$ enters Eq.~\eqref{eq:sdfssdfsdfdsssdfsdfsdffdsfsfd} due to the periodicity of the lattice, which repeats itself with identical rungs existing after every spacing $d$ along the ladder [cf. Fig.~\ref{wfs}~(a)]. Substitution of Eq.~\eqref{eq:sdfssdfsdfdsssdfsdfsdffdsfsfd} into the total Hamiltonian operator $\hat{H}$ of Eq.~\eqref{eq:Haxcsdfdsfdsfsdfvcxvmy} leads to the $q$-space form of the Hamiltonian
\begin{equation}
\label{eq:sdfsdsssdfdsfsdfsdfsdffdsfsfd}
\hat{H} = \sum_{q} \omega_q^A a_{q}^\dagger a_q + \sum_{q} \omega_q^B b_{q}^\dagger b_q + \sum_{q} g \left( a_{q}^\dagger b_q + b_{q}^\dagger a_q \right),
\end{equation}
where the two eigenfrequencies $\omega_{q}^A$ and $\omega_{q}^B$, which are associated with modes belonging to the individual chains A and B respectively, read
\begin{align}
\label{eq:sdfsdssdfgdfgfdsfsfd}
\omega_{q}^A &= \omega_0 + 2 J_1 \cos \left( qd \right) + 2 J_2 \cos \left( 2 qd \right), \\
\omega_{q}^B &= \omega_0 + 2 J_1 \cos \left( qd \right), 
\end{align}
which immediately highlights an important coupling asymmetry amongst the two subsystems. Rearranging the transformed Eq.~\eqref{eq:sdfsdsssdfdsfsdfsdfsdffdsfsfd} into the neater form
\begin{equation}
\label{eq:bfdbfbdfv}
\hat{H} = \sum_{q} \Psi_q^\dagger \mathcal{H}_q \Psi_q,
\end{equation}
 where the $2 \times 2$ Bloch Hamiltonian $\mathcal{H}_q$ and the column vector $\Psi_q$ (which contains the relevant operators) are together defined by
\begin{equation}
\label{eq:sdfsdsssjjijodfdsfsdfsdfsdffdsfsfd}
\mathcal{H}_q = \begin{pmatrix}
\omega_{q}^A & g \\
g & \omega_{q}^B
\end{pmatrix},
\quad\quad\quad\quad\quad
\Psi_q = \begin{pmatrix}
a_q \\
b_q
\end{pmatrix},
\end{equation}
reveals some important properties of the model. Namely, the model respects only one of three important Bloch Hamiltonian symmetries (time-reversal, inversion and chiral) as follows
\begin{align}
\mathcal{H}_q &= \mathcal{H}_{-q}^\ast,  &&\text{(time-reversal symmetry is observed)} \label{eq:dsdsfdsfffs} \\
\mathcal{I} \mathcal{H}_q  \mathcal{I}^{-1} &\ne \mathcal{H}_{-q}, &&\text{(inversion symmetry is not observed)}  \label{eq:sdfsdfsdfsdf} \\
\mathcal{C} \mathcal{H}_q  \mathcal{C}^{-1} &\ne -\mathcal{H}_q, &&\text{(chiral symmetry is not observed)}  \label{eq:ssdfdsfdfsdf} 
\end{align}
where the the inversion operator $\mathcal{I} = \sigma_x$ and the chiral operator $\mathcal{C} = \sigma_z$. Both operators are given in terms of one of the three Pauli spin matrices $\boldsymbol{\sigma} = (\sigma_x, \sigma_y, \sigma_z)$, while $I$ is the identity matrix. As usual, time-reversal symmetry is respected since no magnetic (or even pseudo-magnetic) field is present [cf. Eq.~\eqref{eq:dsdsfdsfffs}]. Inversion symmetry is in general broken because as soon as the the next-nearest-neighbour coupling $J_2 \ne 0$, then the diagonal terms in the Bloch Hamiltonian $\mathcal{H}_q$ are unequal ($\omega_{q}^A \ne \omega_{q}^B$), such that the system is indeed altered upon interchanging type A oscillators with type B oscillators throughout the ladder [cf. Eq.~\eqref{eq:sdfsdfsdfsdf}]. Chiral (or sublattice) symmetry is not obeyed due to the nonzero couplings terms $J_1 \ne 0$ and $J_2 \ne 0$ in the on-diagonal terms of the Bloch Hamiltonian $\mathcal{H}_q$, which ensure that the spectrum is not symmetric about $\omega_0$ (the trivial breaking of chiral symmetry by the constant energy term $\omega_0$, appearing in both $\omega_{q}^A$ and $\omega_{q}^B$, is not consequential as it is just an arbitrary energy shift) [cf. Eq.~\eqref{eq:ssdfdsfdfsdf}]. This brief symmetry analysis suggests that the proposed model likely cannot be classified with a traditional topological index as per the Altland–Zirnbauer table~\cite{Altland1997, Velasco2019}, such that any edge states arising in the two-leg ladder should probably be deemed to be topologically trivial. Indeed, the winding number suggested by the curve formed in a parametric plot of the Bloch Hamiltonian $\mathcal{H}_q = \mathcal{H}_{I} I + \boldsymbol{\sigma} \cdot \boldsymbol{\mathcal{H}}$ over the first Brillouin zone is zero. Here the three-dimensional Hamiltonian vector $\boldsymbol{\mathcal{H}} = (\mathcal{H}_x, \mathcal{H}_y, \mathcal{H}_z)$, and the employed Pauli matrix decomposition implies that the on-diagonal term $\mathcal{H}_{I} = \omega_0 + 2 J_1 \cos \left( qd \right) + J_2 \cos \left( 2 qd \right)$, the off-diagonal term $\mathcal{H}_x = g$ is a constant, the leading diagonal imbalance term $\mathcal{H}_z = J_2 \cos \left( 2 qd \right)$, and finally $\mathcal{H}_y = 0$.

The eigenvalues of the Bloch Hamiltonian $\mathcal{H}_q$ [cf. Eq.~\eqref{eq:sdfsdsssjjijodfdsfsdfsdfsdffdsfsfd}] suggest the pair of eigenfrequencies $\omega_{q ,\tau}$ of the two-leg ladder, which are grouped into two bands (as codified by the index $\tau = \pm 1$) as follows
\begin{equation}
\label{eq:sdfsdsssjjijodfuhyiuyidsfsdfsdfsdffdsfsfd}
\omega_{q ,\tau} = \bar{\omega}_q  + \tau \Omega_q.
\end{equation}
The expression of Eq.~\eqref{eq:sdfsdsssjjijodfuhyiuyidsfsdfsdfsdffdsfsfd} makes use of $\bar{\omega}_q$, the average eigenfrequency of the uncoupled chains A and B encompassing the ladder, and the effective coupling constant $\Omega_q$, which are together defined by
\begin{align}
\bar{\omega}_q &= \frac{ \omega_q^A + \omega_q^B }{2} = \omega_q^B + J_2 \cos \left( 2 qd \right), \label{eq:csdcrv} \\
\Omega_q &= \sqrt{ \Delta_q^2 + g^2 } = \sqrt{ J_2^2 \cos^2 \left( 2 qd \right) + g^2 },  \label{eq:csdcrv2} \\
\Delta_q &=  \frac{ \omega_q^A - \omega_q^B }{2} = J_2 \cos \left( 2 qd \right), \label{eq:csdcr3}
\end{align}
where $\Delta_q$ is the detuning across the two subsystem chains A and B. The eigendecomposition of Eq.~\eqref{eq:sdfsdsssjjijodfdsfsdfsdfsdffdsfsfd}, aided by the eigenfrequencies $\omega_{q ,\tau}$ of Eq.~\eqref{eq:sdfsdsssjjijodfuhyiuyidsfsdfsdfsdffdsfsfd}, directly leads to the diagonal form of the original two-leg ladder Hamiltonian operator $\hat{H}$ of Eq.~\eqref{eq:Haxcsdfdsfdsfsdfvcxvmy} like so
\begin{equation}
\label{eq:sdfsdssdfgdfgsjjijodfdsfsdfsdfsdffdsfsfd}
\hat{H} = \sum_{q, \tau} \omega_{q, \tau} \beta_{q, \tau}^\dagger \beta_{q , \tau},
\end{equation}
where the twin Bogoliubov operators $\beta_{q , \tau}$ introduced above are defined by the operator superpositions
\begin{equation}
\label{eq:sdfsadasda}
\beta_{q, +} = \sin \theta_q  ~ a_q - \cos \theta_q ~ b_q, 
\quad\quad\quad\quad
\beta_{q, -} = \cos \theta_q  ~ a_q + \sin \theta_q ~ b_q, 
\end{equation}
with the two trigonometric Bogoliubov coefficients which have emerged are given by
\begin{equation}
\label{eq:sfdsf}
\cos \theta_q = \tfrac{1}{\sqrt{2}} \left( 1 + \tfrac{\Delta_q}{\Omega_q} \right)^{\frac{1}{2}}, 
\quad\quad\quad\quad\quad
\sin \theta_q = \tfrac{1}{\sqrt{2}} \left( 1 - \tfrac{\Delta_q}{\Omega_q} \right)^{\frac{1}{2}}, 
\end{equation}
which are determinable from the two quantities $\Omega_q$ and $\Delta_q$, as defined in Eq.~\eqref{eq:csdcrv2} and Eq.~\eqref{eq:csdcr3} respectively. The knowledge of the eigenfrequencies and eigenstates of the two-leg ladder allows for some general properties of the system to be inferred within standard band theory, and we restrict our analysis to the first Brillouin zone $-\pi/d < q \le \pi / d$ only due to the periodicity of the considered lattice geometry.

The group velocity $v_{q, \tau} = \partial_q \omega_{q, \tau}$ for a mode in the band $\tau = \pm 1$ and with the quantum number $q$, follows directly from Eq.~\eqref{eq:sdfsdsssjjijodfuhyiuyidsfsdfsdfsdffdsfsfd} as
\begin{equation}
\label{eq:sfdsdf}
v_{q, \tau} = - 2 d \left[ J_1 \sin \left( qd \right) + J_2 \sin \left( 2 q d \right) \right] -\tau \frac{J_2^2 d}{\Omega_q} \sin \left( 4 q d \right).
\end{equation}
Clearly there are always, that is for both bands $\tau = \pm 1$ and for any values of $J_2/J_1$ and $g/J_1$, roots of the group velocity $v_{q, \tau}$ for modes at the centre and edge of the Brillouin zone with the quantum numbers $q = 0$ and $q = \pi /d$ respectively. For larger coupling ratios $J_2/J_1$, these two guaranteed roots are joined by additional stationary points. For example, in the extreme limit of $g \gg J_2$ the group velocity $v_{q, \tau}$ reduces to the band index independent quantity $v_{q, \tau} = - 2 d \left[ J_1 \sin \left( qd \right) + J_2 \sin \left( 2 q d \right) \right]$. Upon transforming the variable from $qd$ to $z = \mathrm{e}^{\mathrm{i} q d}$, the roots of the group velocity can be then be found by solving the quartic equation $J_2 z^4 + J_1 z^3 - J_1 z - J_2 = 0$. The already known roots of $q = 0$ and $q = \pi /d$ correspond to the twin solutions of $z  = \pm 1$, essentially reducing the problem to the effectively quadratic equation $(z+1)(z-1)(J_2 z^2 + J_1 z + J_2) = 0$. The final two roots are then found to be located at $z = -J_1/(2J_2) \pm \mathrm{i} \sqrt{1 - J_1^2/(2J_2)^2}$, or equivalently at $q d = \pm ( \pi - \arctan \sqrt{(2J_2/J_1)^2-1})$. This brief analysis suggests additional stationary points only occur above a critical coupling ratio $J_2 / J_1 > 1/2$, at least in the considered regime $g \gg J_2$. A similar calculation may be performed in the opposing limit of $g \ll J_2$, when the two subsystems effectively decouple, which reveals a different threshold coupling ratio of $J_2 / J_1 > 1/4$ for additional stationary points to emerge (see App.~\ref{Sec:chainA} for the derivation). These additional stationary points in the bandstructure are found out to be highly consequential for the emergence of edge states, as is discussed later on.

The bandgap $\delta \omega$, defined as the energy difference between the minimum of the upper band $\omega_{q, +}$ and the maximum of the lower band $\omega_{q, -}$ for any value of the quantum number $q$, is given by the intuitive formula
\begin{equation}
\label{eq:sfdfdgdggnnsdf}
\delta \omega = \mathrm{min} \{ \omega_{q, +} \} - \mathrm{max} \{ \omega_{q, -} \},
\end{equation}
which may be positive, negative or zero in the case of the considered ladder model. In particular, a negative bandgap $\delta \omega$ suggests band overlap since the lower band will necessarily protrude into the territory of the nominally upper band. In the simplest case of very weak next-nearest neighbour coupling $J_2 \ll J_1$, a negative bandgap only exists when the inequality $g < 2 J_1$ holds, since $\mathrm{min} \{ \omega_{q, +} \} = \omega_0 - 2 J_1 + g$ and $\mathrm{max} \{ \omega_{q, -} \} = \omega_0 + 2 J_1 - g$ in this limit, such that the bandgap of Eq.~\eqref{eq:sfdfdgdggnnsdf} becomes $\delta \omega = 2 (g - 2 J_1)$ within this simple regime. The influence of the bandgap $\delta \omega$ on the appearance of edge states is discussed in detail in the next section.

\begin{figure*}[tb]
 \includegraphics[width=0.97\linewidth]{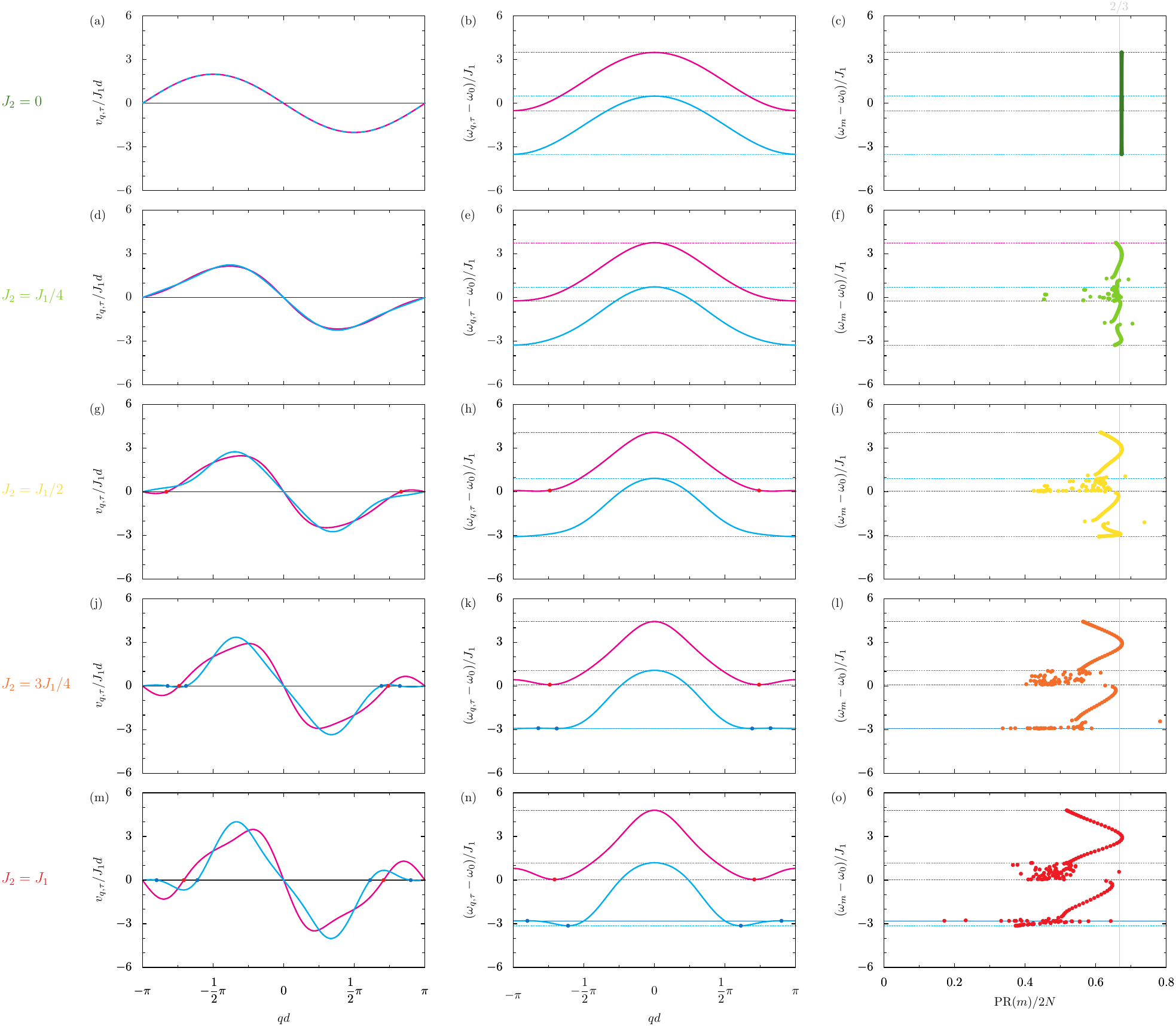}
 \caption{ \textbf{Stationary points, bandstructure and degrees of localization in the two-leg ladder model.} We consider five increasingly large values of the next-nearest neighbour coupling $J_2$ upon descending the columns of the figure (as marked on the left-hand side), and we set the inter-chain coupling $g = 3 J_1/2$. First column: the group velocity $v_{q, \tau}$ as a function of the quantum number $q$, in the first Brillouin zone $-\pi/d < q \le \pi / d$. The results for the $\tau = +1$ (cyan lines) and $\tau = -1$ (pink lines) bands are shown. The calculation is carried out with periodic boundary conditions for a two-leg ladder in the limit of $2 N \to \infty$ oscillators [cf. Eq.~\eqref{eq:sfdsdf}]. In panels (g, j, m), the significant $J_2$ couplings have given rise to extra roots in the group velocity (marked by pink and cyan circles), corresponding to additional stationary points in the bandstructure $\omega_{q, \tau}$, which are similarly marked in panels (h, k, n). Second column: the eigenfrequencies $\omega_{q, \tau}$ forming the periodic bandstructure [cf. Eq.~\eqref{eq:sdfsdsssjjijodfuhyiuyidsfsdfsdfsdffdsfsfd}] and corresponding to the results of the first column. Dashed pink and cyan horizontal lines: the extrema of each ($\tau = \pm 1$) band, which are similarly carried through to the third column, and which imply a series of negative bandgaps [cf. Eq.~\eqref{eq:sfdfdgdggnnsdf}]. Solid blue horizontal lines in panels (k) and (n): the modes associated with additional stationary points in the lower $\tau = - 1$ band, which are carried through to panels (l) and (o) in the third column. Third column: the scaled participation ratio $\mathrm{PR} \left( m \right) / 2 N$ of each state $m$ [cf. Eq.~\eqref{eq:sdfddwsdfs}], which reside at the eigenfrequency $\omega_{m}$. Solid grey verticle lines: guide for the eye at $2/3$. The numerical calculation in the third column is carried out with open boundary conditions for a two-leg ladder with $2N = 200$ oscillators.}
 \label{tony}
\end{figure*}

We present graphically some visualizations of a few key results coming from the bandstructure of the two-leg ladder in the first two columns of Fig.~\ref{tony}, for the example case with the inter-chain coupling $g = 3 J_1/2$. The first column of panels (a, d, g, j, m) in Fig.~\ref{tony} display the group velocity $v_{q, \tau}$ in the first Brillouin zone using Eq.~\eqref{eq:sfdsdf}, where descending the column increases the next-nearest neighbour coupling strength $J_2$  (as marked on the left-hand side of Fig.~\ref{tony}). The results for the upper band $\tau = +1$ are denoted with magenta lines, while the lower $\tau = -1$ band are associated with cyan lines. As discussed after Eq.~\eqref{eq:sfdsdf}, the guaranteed stationary points at $q = 0$ and $q = \pi /d$ are joined by additional stationary points for the cases of larger next-nearest neighbour couplings $J_2$, as marked by pink circles and cyan circles in the relevant Fig.~\ref{tony}~(g, j, m), where $J_2 \ge J_1/2$. The second column of Fig.~\ref{tony} shows the eigenfrequencies $\omega_{q, \tau}$ corresponding to the cases of the first column and forming the exact bandstructure of the two-leg ladder via Eq.~\eqref{eq:sdfsdsssjjijodfuhyiuyidsfsdfsdfsdffdsfsfd}. The dashed pink and cyan lines mark the extrema of each ($\tau = \pm 1$) band, which imply a series of negative bandgaps [cf. Eq.~\eqref{eq:sfdfdgdggnnsdf}] within the chosen parameter regime. Importantly, the final three panels (h, k, n) in Fig.~\ref{tony} contain pink circles and cyan circles in the same manner as the group velocity panels (g, j, m), which correspond to stationary points in the bandstructure. In particular, the solid blue lines in panels (k) and (n) mark the frequencies $\omega_{q, \tau}$ of the modes which are associated with additional stationary points in the lower $\tau = - 1$ band, which lies inside the energetic sector created by the upper and lower bounds of the lower $\tau = - 1$ band (as denoted by the two dashed cyan lines). As such, if these modes are edge states, and if they are surrounded by a collection of extended states comprising the rest of the effective band, the highlighted modes will in essence be a type of bound state in the continuum~\cite{Hsu2016, Azzam2020, Sadreev2021, Koshelev2023}. In the next Sec.~\ref{Sec:Finitesystems}, we finally consider the proper finite-sized two-leg ladder system as defined by Eq.~\eqref{eq:Haxcsdfdsfdsfsdfvcxvmy}, which allows us to probe the existence of edge states within the lattice model and their relation to certain bulk properties.
\\


\section{Finite-sized systems}
\label{Sec:Finitesystems}

Let us now consider the two-leg ladder Hamiltonian $\hat{H}$ as originally introduced in Eq.~\eqref{eq:Haxcsdfdsfdsfsdfvcxvmy} for a ladder of finite length $N$, such that there are $2N$ coupled oscillators in the overall system. Upon staying in real space [instead of moving into $q$-space using the transformation of Eq.~\eqref{eq:sdfssdfsdfdsssdfsdfsdffdsfsfd}] the Hamiltonian matrix $\mathcal{H}$ will be of size $2N \times 2N$ [instead of the $2 \times 2$ Bloch Hamiltonian matrix $\mathcal{H}_q$ of Eq.~\eqref{eq:sdfsdsssjjijodfdsfsdfsdfsdffdsfsfd}, as found within $q$-space]. The diagonalization of the finite square matrix $\mathcal{H}$ (with implied open boundary conditions) is readily achieved, leading to $2N$ eigenfrequencies $\omega_m$, where the integer index $m = \{ 1, 2, ..., 2N \}$. For each state labelled by $m$, the $n$-th element of the associated eigenvector can be written as $c_n (m)$, and there will be $2 N$ elements in total, corresponding to each site in the two-leg ladder. The knowledge of the eigenvectors allows for the spread of the state across the whole lattice to be quantified using a popular localization measure: the participation ratio $\mathrm{PR} \left( m \right)$ of a state $m$. This quantity may be defined via the formula~\cite{Bell1970, Thouless1972}
\begin{equation}
\label{eq:sdfddwsdfs}
\mathrm{PR} \left( m \right) = \frac{ \left( \sum_{n} | c_n (m) |^2 \right)^2}{\sum_{n} | c_n (m) |^4},
\end{equation}
where the numerator is simply a normalization, while the denominator discriminates between states with different spatial extents. In the case of the two-leg ladder considered here, the sums in Eq.~\eqref{eq:sdfddwsdfs} are taken over the whole array $n  = \{ 1, 2, ..., 2N \}$. A fully localized state is trapped on just one site $n$, so that $\mathrm{PR} \left( m_\text{f.~localized} \right) = 1$. Conversely, a fully extended state will feel all $2N$ sites equally, such that its participation ratio will scale exactly with the system size as $\mathrm{PR} \left( m_\text{f.~extended} \right) = 2N$. Otherwise, typical extended states will observe a linear with system size relationship like $\mathrm{PR} \left( m_\text{extended} \right) \simeq \beta (2N)$, where the prefactor $\beta$ is some dimensionless number which is known exactly for certain simple cases (see Refs.~\cite{Torres2019, Schiulaz2018} for example, while App.~\ref{Sec:chainB} contains a derivation of the common prefactor $\beta = 2/3$ for a single regular chain). Away from these common scenarios, that is those with participation ratios scaling like $(2N)^0$ or $(2N)^1$, states exhibiting the more peculiar expression $\mathrm{PR} \left( m \right) \simeq \beta (2N)^\alpha$ are of great interest. Here the exponent $\alpha$ satisfies the inequality $0 < \alpha < 1$, such that the state has a distinctly sublinear dependence on the overall system size $2N$. These unusual states are then arguably localized to some degree, depending upon the exact size of the key exponent $\alpha$. In some contexts (including quasiperiodic models), these states are sometimes known as multifractal states, where $\alpha$ is related to the fractal dimension of the state~\cite{Sarkar2021, Duthie2022, Aditya2023}. 

We explore the various degrees of localization of the states supported in the two-leg ladder in the third column of Fig.~\ref{tony}. As in the first and second columns of Fig.~\ref{tony}, the inter-chain coupling remains at $g = 3 J_1/2$ and going down the column of panels~(c, f, i, l, o) increases the next-nearest neighbour coupling $J_2$, but now we consider a finite-sized array of $2N = 200$ oscillators as described by Eq.~\eqref{eq:Haxcsdfdsfdsfsdfvcxvmy}. We plot in Fig.~\ref{tony}~(c, f, i, l, o) the eigenfrequencies $\omega_m$, where each state is labelled by the index $m$, as a function of its participation ratio $\mathrm{PR} \left( m \right)$ [cf. Eq.~\eqref{eq:sdfddwsdfs}]. The simplest case of zero next-nearest neighbour coupling ($J_2 = 0$) is displayed in Fig.~\ref{tony}~(c), where the common participation ratio of $\mathrm{PR} \left( m \right) \simeq (2/3) (2N)$ for all states $m$ is most apparent, signalling rather standard extended states (indeed, the prefactor of $\beta \simeq 2/3$ is obtained analytically in App.~\ref{Sec:chainB}). In Fig.~\ref{tony}~(f, i), where the next-nearest neighbour coupling $J_2$ is nonzero but still relatively small, there are fluctuations around the value of $2/3$ (a number marked with a solid grey vertical line throughout the third column) suggesting typical extended states but without the complete uniformity of panel~(c). Notably especially in panel (i), where $J_2 = J_1/2$, two collections of states have started to accumulate around the band edges of the upper and lower band (as found from the continuum calculation, and as marked with the dashed magenta and cyan horizontal lines in each panel). The cases with the largest next-nearest neighbour coupling $J_2$ in the lowest two panels finally reveal the formation of significantly less extended states, with the smallest scaled participation ratio being $\mathrm{min} \{ \mathrm{PR} \left( m \right) \} / 2 N \simeq 0.171$ in Fig.~\ref{tony}~(o), where $J_2 = J_1$. This state of minimum participation ratio in panel (o) resides at a certain eigenfrequency $\omega_m$, as marked by a solid blue line, which corresponds to an eigenfrequency $\omega_{q, \tau}$ coming from the continuum model. This particular eigenfrequency $\omega_{q, \tau}$ is shown in the adjacent panel (n) by the same solid blue line. The specific quantum number $q$ associated with this particular eigenfrequency $\omega_{q, \tau}$ is distinguished by a blue circle in panel (n), since it is a stationary point [all additional stationary points are marked with circles in the first two columns of Fig.~\ref{tony}, as discussed after Eq.~\eqref{eq:sfdsdf} and as confirmed in the group velocity plots given in panels (g, j, m)]. Notably, the most localized state in panel (o) is also seen to be housed inside the notional spectral band of extended states (limited by the dashed horizontal lines), a defining characteristic of a bound state in the continuum~\cite{Hsu2016, Azzam2020, Sadreev2021, Koshelev2023}.

We visualize two characteristic states in the two-leg ladder in Fig.~\ref{wfs}~(b, c), where the couplings $g = 3 J_1/2$ and $J_2 = J_1$, and we consider a system with $2N = 200$ oscillators [which matches the parameter situation in Fig.~\ref{tony}~(o)]. We plot the probability density over the $2N$ sites of the array, where the results for the $A$ and $B$ sites $n$ are shown with pink and yellow lines respectively. Notably, the example extended state shown in  in Fig.~\ref{wfs}~(b) feels around two-thirds of the array, its scaled participation ratio $\mathrm{PR} \left( m \right) / 2 N \simeq 0.675$ and its oscillatory structure is noticeably balanced across both subsystems chain A and chain B. In Fig.~\ref{wfs}~(c) we finally reveal the spatial structure of the more localized state discussed in Fig.~\ref{tony}~(o), of scaled participation ratio $\mathrm{PR} \left( m \right) / 2 N \simeq 0.171$, which is seen to be a type of edge state due to its significant probability density at the ends of the ladder (across both chain A and chain B). The correspondence between the appearance of an edge state in the finite system and the emergence of additional stationary points in the continuum bandstructure is somewhat reminiscent of the celebrated bulk-edge correspondence in topological matter (where a calculation of a topological invariant from a continuum theory predicts the presence or absence of topological edge states in the finite version of the model~\cite{Hasan2010, Asboth2016}).

\begin{figure*}[tb]
 \includegraphics[width=\linewidth]{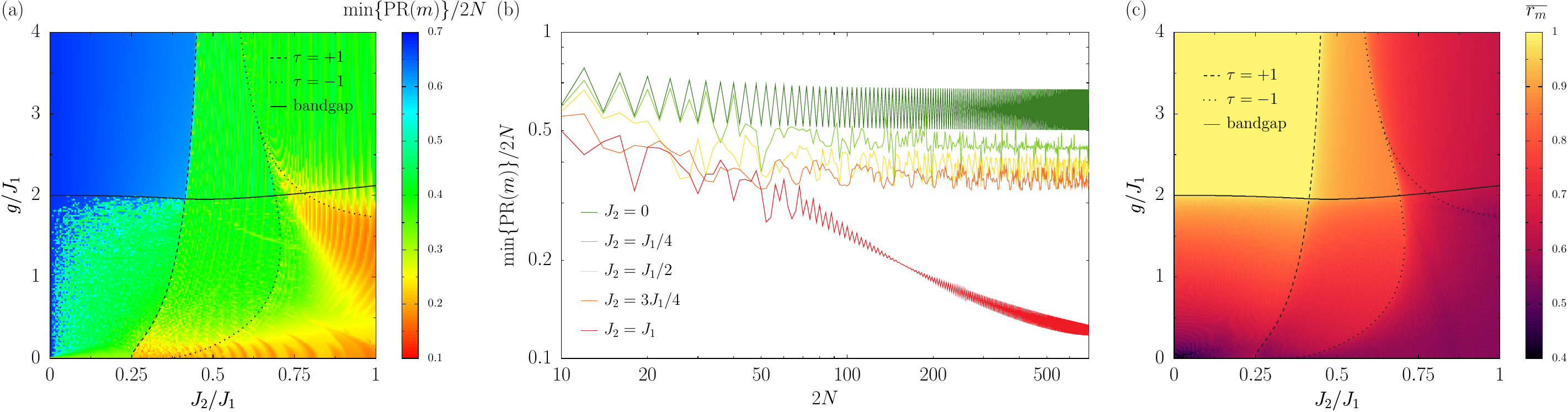}
 \caption{ \textbf{Degrees of localization and energy level statistics in the two-leg ladder model.} Panel (a): a map of the minimum of the scaled participation ratio $\mathrm{min} \{ \mathrm{PR} \left( m \right) \} / 2 N$, as a function of both the inter-chain coupling strength $g$ and the next-nearest neighbour coupling strength $J_2$, both in units of the nearest-neighbour coupling $J_1$ [cf. Eq.~\eqref{eq:sdfddwsdfs}]. Dashed black line: boundary marking areas with different numbers of stationary points for the upper $\tau = +1$ band. Dotted black line: boundary marking areas with different numbers of stationary points for the lower $\tau = -1$ band. Solid black line: the borderline below which there is a negative bandgap. These indicative black lines are calculated from the continuum results of Eq.~\eqref{eq:sdfsdsssjjijodfuhyiuyidsfsdfsdfsdffdsfsfd}, Eq.~\eqref{eq:sfdsdf} and Eq.~\eqref{eq:sfdfdgdggnnsdf}. In this panel, we consider a two-leg ladder with $2N = 200$ oscillators. Panel (b): we take a horizontal cut through panel (a) at $g = 3J_1/2$, and then we make vertical cuts at five different values of $J_2$ (as noted in the plot legend), before extending the results as a function of the system size $2N$ in a log–log plot (up to $2N = 700$). Panel (c): a map of the mean consecutive energy level spacings $\overline{r_m}$, as a function of both $g$ and $J_2$ [cf. Eq.~\eqref{eq:dfsf}]. In this panel, we consider a two-leg ladder with $2N = 2000$ oscillators and the indicative black lines are the same as the lines provided in panel (a).}
 \label{numer}
\end{figure*}

We chart the degrees of localization of states supported by the two-leg ladder system more generally in Fig.~\ref{numer}. As a function of both the next-nearest neighbour coupling ratio $J_2/J_1$ and the inter-chain coupling ratio $g/J_1$, we plot the minimum of the scaled participation ratio $\mathrm{min} \{ \mathrm{PR} \left( m \right) \} / 2 N$ in Fig.~\ref{numer}~(a), where we consider a finite system again composed of $2N = 200$ oscillators. Blue and green regions in panel~(a) represent cases where the ladder only supports highly extended states, which display the typical behaviour of $\mathrm{PR} \left( m \right) / 2 N \sim 2/3$, while more interestingly the orange-red regions in the map suggest that the system may host somewhat localized states, since the localization measure $\mathrm{PR} \left( m \right) / 2 N \sim 1/10$. The map of Fig.~\ref{numer}~(a) demonstrates the interplay between the key ratios $J_2/J_1$ and $g/J_1$: the next-nearest neighbour coupling ratio $J_2/J_1$ needs to be sufficiently large (in this case, $J_2 > J_1/4$) for a red phase to first appear, while if the inter-chain coupling ratio $g/J_1$ is too large (roughly, $g > 2 J_1$) no red phase is supported. The general character of the diagram of Fig.~\ref{numer}~(a) can be understood using some key results from the band theory of the continuum version of the model. Explicitly, as was discussed after Eq.~\eqref{eq:sfdsdf} and also after Eq.~\eqref{eq:sfdfdgdggnnsdf}, some limiting cases of the properties of the two-leg ladder with periodic boundary conditions suggest that
\begin{align}
  &\text{additional stationary points arise when} \quad &&J_2 > \frac{J_1}{4},   &&&\text{for} \quad g \ll J_2, \label{eq:sfdsdfssdfsfdsdf} \\
  &\text{additional stationary points arise when}  &&J_2 > \frac{J_1}{2},  &&&\text{for} \quad g \gg J_2, \label{eq:sfdsdfssdfsfdsdf2} \\
  &\text{a negative bandgap arises when}  &&g < 2 J_1, &&&\text{for} \quad J_2 \ll J_1. \label{eq:sfdsdfssdfsfdsdf3}
\end{align}
The inequality of Eq.~\eqref{eq:sfdsdfssdfsfdsdf} explains the critical behaviour along the bottom horizontal axis of Fig.~\ref{numer}~(a), as marked by the emergence of the red-orange colours, corresponding to the first appearance of stationary points due to next-nearest neighbour interactions. Meanwhile Eq.~\eqref{eq:sfdsdfssdfsfdsdf2} is suggested at the top horizontal axis of Fig.~\ref{numer}~(a), due to the transition from blue to green regions. The band overlap arising from a negative bandgap $\delta \omega$ arises when Eq.~\eqref{eq:sfdsdfssdfsfdsdf3} is true, which actually approximately holds rather nicely for all of the range $0 \le J_2 \le J_1$. The solid black line in Fig.~\ref{numer}~(a) demarcates the exact line in parameter space below which the bandgap $\delta\omega$ is negative [cf. Eq.~\eqref{eq:sfdfdgdggnnsdf}]. Similarly, the dashed black line marks areas with different numbers of stationary points for the upper band ($\tau = +1$), while the dotted black line does the same job for the lower band ($\tau = -1$). These three guides for the eye, arising from purely bulk calculations, certainly seem to coincide with interesting behaviour of the corresponding finite system with edges. As an aside, when tending towards zero inter-chain coupling ($g \to 0$) we find a singular limit: just above the horizontal axis of Fig.~\ref{numer}~(a) the red-orange colours when $J_2 > J_1/4$ refer to significantly smaller participation ratios, but this behaviour is not found exactly at $g = 0$ (see for example the analyses of the individual chain A and chain B models in App.~\ref{Sec:chainA} and App.~\ref{Sec:chainB} respectively). The occurrence of this singular limit lends some support to the notion that it is important for greater degrees of localization that the system contains subsystems with different boundary conditions, as here the upper and lower chain subsystems have contrasting boundary conditions due to their different coupling ranges (which is only not relevant exactly when $g = 0$, as then the subsystems are completely decoupled).

A more accurate categorization of the nature of the extended states and edge states supported by the two-leg ladder requires an analysis which tracks changes due to size-dependencies. In Fig.~\ref{numer}~(b) we effectively take a horizontal cut through panel~(a) at the specific inter-chain coupling $g = 3 J_1/2$,  and then we make vertical cuts at five different values of $J_2$ (as noted in the plot legend). We then extending the result, the minimum of the scaled participation ratio $\mathrm{min} \{ \mathrm{PR} \left( m \right) \} / 2 N$, as a function of the total number of oscillators in the ladder $2N$. This process results in the log–log plot presented as Fig.~\ref{numer}~(b). Four of the lines in Fig.~\ref{numer}~(b) effectively tend towards a constant (or oscillate between two constants in the case of the dark green line, where $J_2 = 0$) which suggests that the participation ratio scales linearly with $2N$: this data truly describes an extended phase of the system. However, the red line in Fig.~\ref{numer}~(b), representing the case of $J_2 = J_1$, does not seem to tend towards a constant for the considered system sizes such that the participation ratio scales sublinearly with $2N$ (at least within this regime up to $2N = 700$). This suggests an unconventional phase containing edge states [cf. the edge state depicted in Fig.~\ref{wfs}~(c) for the case of $2N = 200$ oscillators], albeit not fully localized states since the relationship is not completely independent of $2N$. Similar plots to Fig.~\ref{numer}~(b) may be found in App.~\ref{Sec:chainAB} for other values of the specific inter-chain coupling $g$, providing further information about the overall phase diagram of Fig.~\ref{numer}~(a). Notably, we draw our conclusions from these aforementioned finite size effect plots based upon system sizes from $\sim 10^1$ to $\sim 10^3$, since this regime can plausibly be explored experimentally~\cite{Periwal2021, ZhuGao2022, Martinez2021}. We leave to a later study the numerical crunching of two-leg ladders containing larger numbers of oscillators, where the already-discovered non-monotonic nature of the data may revise some of the scalings between some ranges of exceedingly large $2N$ (in which case, the potential consequences will only effect the experimental study of extremely large systems). We also briefly discuss in App.~\ref{Sec:detune} the impact of a nonzero energy detuning between the resonance frequencies in the upper chain and lower chain as a complement to the data presented in Fig.~\ref{numer}, an analysis which also may be useful to guide certain future experiments and which hints at the possibility for inducing stronger degrees of localization thanks to energy detuning.

We further characterize the behaviour of the two-leg ladder by considering the energy level statistics of the model~\cite{Mehta2014, Guhr1998}. For a generic $2N \times 2N$ matrix Hamiltonian, we may list the resulting eigenvalues $E_m$ in ascending order, where the index $m = \{ 1, 2, ..., 2N \}$. Therefore, there are $2N-1$ consecutive energy level spacings $S_m$, where $S_m = E_{m+1}-E_m$. The ratio of two consecutive gaps $r_m$ may then be defined as~\cite{Oganesyan2007, Atas2013}
\begin{equation}
\label{eq:dfsf}
r_m = \frac{\mathrm{min} \{ S_{m},S_{m-1} \}}{\mathrm{max} \{ S_{m}, S_{m-1} \}},
\end{equation}
which observes the bounds $0 \le r_m \le 1$, where the index runs for $m = \{  2, 3, ..., 2N-1 \}$. The dimensionless quantity $r_m$ measures the correlations between the adjacent energy level gaps in a given spectrum. The mean value $\overline{r_m}$, found after averaging over all $2N-1$ ratios $r_m$, acts as a kind of judge of the chaoticity appearing in the system. Some typical values of $\overline{r_m}$ for different categories of energy-level statistics are provided in Refs.~\cite{Torres2019, Atas2013}: Poissonian level statistics are associated with $\overline{r_m} = 2 \ln 2 -1 \simeq 0.39$, while Gaussian orthogonal ensembles are characterized with $\overline{r_m} = 4 - 2 \sqrt{3} \simeq 0.54$. In ordered (picket fence-like) models $\overline{r_m}$ should essentially be unity. Localization is implied with Poissonian-like values of $\overline{r_m}$, since localized states can reside in different single-particle basis states of nevertheless similar energies, such that they will not interact or display any significant level repulsion~\cite{Oganesyan2007}. In the case of the considered two-leg ladder model, we associate the eigenfrequencies $\omega_m$ with the energy levels $E_m$ alluded to above, and we apply averaging over Eq.~\eqref{eq:dfsf} in order to generate the results presented in Fig.~\ref{numer}~(c). In the map of panel~(c), we plot $\overline{r_m}$ as function of both the next-nearest neighbour coupling ratio $J_2/J_1$ and the inter-chain coupling ratio $g/J_1$, while the black lines acting as guides for the eye are exactly those of Fig.~\ref{numer}~(a). Notably, the yellow region of Fig.~\ref{numer}~(c) corresponds to an ordered system with $\overline{r_m} \simeq 1$, implying no collections of localized states. Meanwhile, different flavours of extended phases are suggested by the orange and red regions at the top of panel~(c), which are approximately encased by the indicative black lines. The lower half of Fig.~\ref{numer}~(c) sees red areas gradually merge into purple islands, in particular the enclosed region ending at $J_2 = J_1$, which are associated with $\overline{r_m} \simeq 0.4$. These kinds of averaged energy level statistics instead suggest a (relatively) more localized phase, as is consistent with the participation ratio map of Fig.~\ref{numer}~(a). Similar conclusions based upon the averaging of the quantity defined in Eq.~\eqref{eq:dfsf} have recently been drawn in both theoretical and experimental studies of a variety of other physical systems hosting collections of relatively localized states~\cite{Cuevas2012, Agarwal2015, Morong2021}.
\\


\section{Discussion}
\label{Sec:Discussion}

In conclusion, we have presented a basic theory for the generation of unconventional edge states in bipartite coupled mode systems due to the presence of asymmetric interactions. We have concentrated on the simplest case of a two-leg ladder model, where we allowed for different coupling ranges in the upper chain as compared to the lower chain. The long-range coupling leads to additional stationary points developing in the bandstructure which, in a remarkable display of a seemingly non-topological bulk-edge correspondence, predicts the existence of states residing at the edges of the analogous finite-sized lattice. Since these edge states are essentially surrounded by a band of extended states they are in some sense a kind of bound state in the continuum~\cite{Hsu2016, Azzam2020, Sadreev2021, Koshelev2023}, albeit generated by the unusual mechanism of asymmetric coupling ranges. Our theoretical predictions could be realized in various physical platforms modelled by lattices with significant next-nearest neighbour interactions, for example with atomic, photonic, phononic and polaritonic setups~\cite{Periwal2021, Martinez2021, ZhuGao2022, Bossart2023, Rajabpoor2023, Caselli2015, Senanian2023, Alyatkin2020, Dovzhenko2023}. Perhaps the acoustic metamaterial platform of Ref.~\cite{ZhuGao2022} is the most promising avenue for immediate experimental realization due to its tunability. Finally, although we have concentrated on a toy model of a two-leg ladder for simplicity, our results should generalize to any bipartite systems with asymmetries in their respective couplings, for example matter modes interacting with radiative modes~\cite{Downing2021, Zhou2022, Allard2022, Allard2023}.
\\


\appendix

\renewcommand{\theequation}{A \arabic{equation}}
\section{The chain A model}
\label{Sec:chainA}


In this appendix we consider the chain A model only, as defined by the Hamiltonian operator $\hat{H}_A$ of Eq.~\eqref{eq:sdfsdfsfd}. It describes a linear chain of quantum harmonic oscillators (each of resonance frequency $\omega_0$) interacting with both nearest-neighbour couplings $J_1$ and -- most importantly -- also next-nearest-neighbour couplings $J_2$ [see the sketch of chain A in Fig.~\ref{wfs}~(a)]. We are especially interested in the continuum limit of a very long chain ($N \to \infty$), and we commonly employ periodic boundary conditions in order to obtain an analytic understanding of both the band theory and the energy level statistics of this subsystem model. In particular, we analyze the additional stationary points in the bandstructure which appear for stronger next-nearest neighbour couplings, we present a detailed derivation of the analytic form of the density of states beyond nearest-neighbour interactions, and we explore in detail the probability density function of adjacent energy levels.

\begin{figure*}[tb]
 \includegraphics[width=\linewidth]{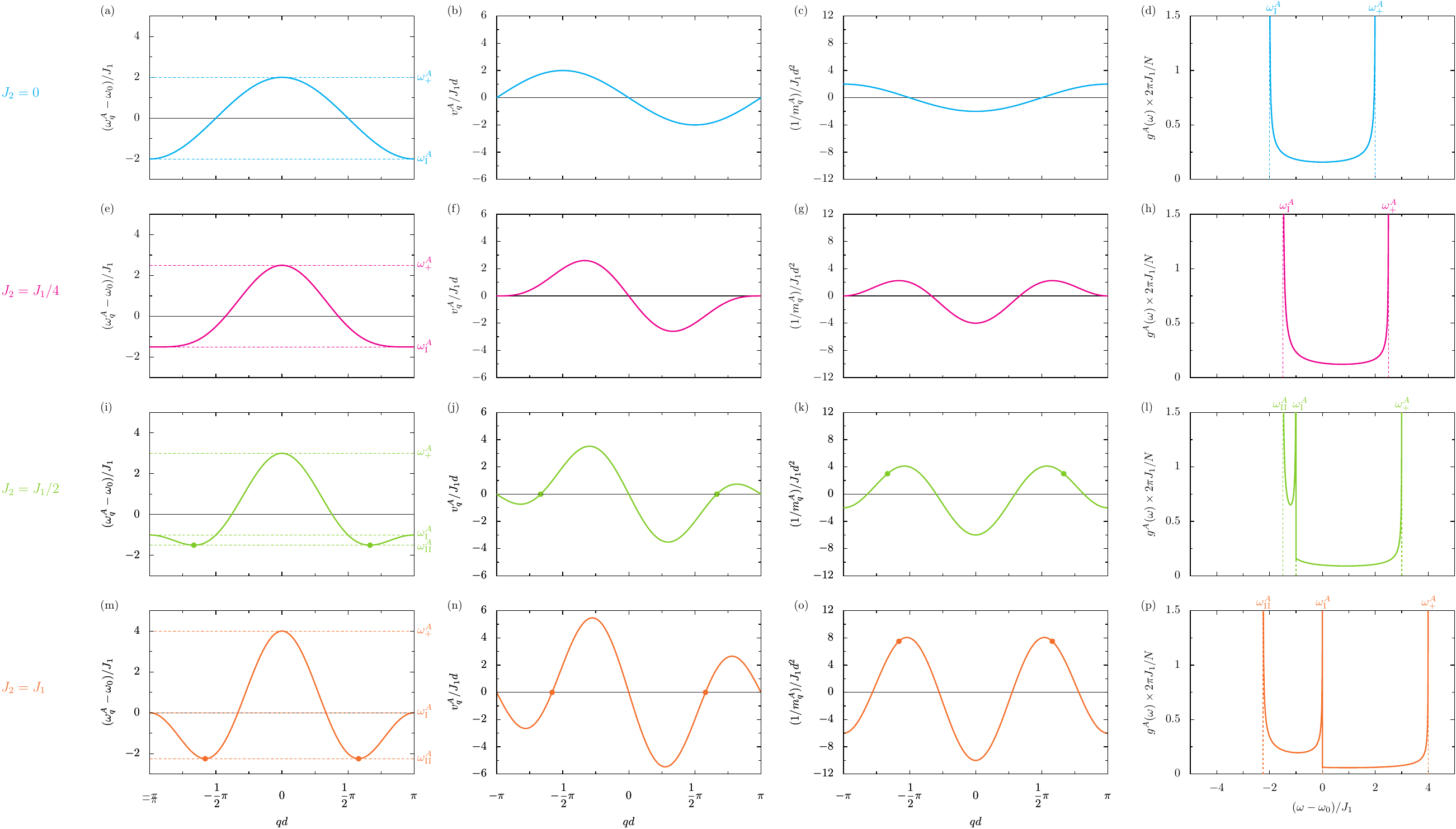}
 \caption{ \textbf{Band properties of the chain A model for an infinite system with periodic boundary conditions.} We consider four increasingly large values of the next-nearest neighbour coupling $J_2$ upon descending the columns of the figure. First column: the bandstructure $\omega_{q}^A$, as a function of the quantum number $q$, in the first Brillouin zone $-\pi/d < q \le \pi / d$ [cf. Eq.~\eqref{eq:gdfdgf2}]. Second column: the associated group velocity $v_{q}^A$ [cf. Eq.~\eqref{eq:sdf}]. Third column: the associated inverse effective mass $1/m_{q}^A$ [cf. Eq.~\eqref{eq:sdf2}]. Fourth column: the density of states $g \left( \omega \right)$, as a function of frequency $\omega$ [cf. Eq.~\eqref{eq:sfvfdfsdffsfdsgrtr} for $0 \le J_2 \le J_1/4$ and Eq.~\eqref{eq:sfvfdfnlsdsdfsdfffsfdsgrtr} for $J_1/4 \le J_2 \le J_1$]. Coloured circles in the bottom two rows: critical values of $q$ associated with additional stationary points of $\omega_{q}^A$ (that is, other than those at $q = 0$ and $q = \pi/d$) [cf. Eq.~\eqref{eq:sdfsdfsdfvdfsdf}]. Dashed lines in the first and fourth columns: the band maximum $\omega_{+}^A$ [cf. Eq.~\eqref{eq:gsdfdgfhfghfsdfdgf}] and the eigenfrequencies $ \omega_{\mathrm{I}}^A$ and $ \omega_{\mathrm{II}}^A$ associated with stationary points [cf. Eq.~\eqref{eq:gsdfdgfhdfgdfgdfgfdgffghfsdfdgf} and Eq.~\eqref{eq:sfdvdffvfe}].}
 \label{chin}
\end{figure*}

The exponential Fourier transform of Eq.~\eqref{eq:sdfssdfsdfdsssdfsdfsdffdsfsfd} immediately diagonalizes the chain A Hamiltonian of Eq.~\eqref{eq:sdfsdfsfd}, which reveals the eigenfrequencies $\omega_{q}^A$ associated with the quantum number $q$ as follows
\begin{align}
\label{eq:gdfdgf}
\hat{H}_A &= \sum_{q} \omega_q^A a_{q}^\dagger a_q, \\
\omega_{q}^A &= \omega_0 + 2 J_1 \cos \left( qd \right) + 2 J_2 \cos \left( 2 qd \right).  \label{eq:gdfdgf2}
\end{align}
The assumed periodic boundary conditions ensure the $2 \pi$-periodicity of the eigensystem, allowing us to focus on the first Brillouin zone $-\pi/d < q \le \pi / d$ only. Essentially, the nonzero inter-site couplings $J_1$ and $J_2$ in the chain A model allow for excitations to move throughout the one-dimensional lattice, which gives rise to kinetic energy terms [cf. the second and third terms in Eq~\eqref{eq:gdfdgf2}] in addition to the onsite energy term of each individual oscillator [cf. the first term in Eq~\eqref{eq:gdfdgf2}]. The energy bands formed from $\omega_{q}^A$ are plotted in the first column of Fig.~\ref{chin} using Eq~\eqref{eq:gdfdgf2}, for four increasingly large values of the next-nearest neighbour coupling $J_2$ upon descending the panels (a, e, i, m). Most notably, with larger next-nearest couplings $J_2$ in Fig.~\ref{chin}~(i, m) the bandstructure exhibits additional stationary points, as marked by coloured circles in the figure.

The extrema $\omega_{\pm}^A$ of the bandstructure defined by Eq~\eqref{eq:gdfdgf2} are also noticeably affected by sufficiently strong $J_2$, since the maximum and minimum eigenfrequencies are given by
\begin{align}
\label{eq:gsdfdgfhfghfsdfdgf}
\omega_+^A &= \omega_0 + 2 \left( J_1 + J_2 \right), \\
\omega_-^A &= \begin{cases}
 \omega_{\mathrm{I}}^A,  & \quad\quad\quad  0 \le J_2 \le J_1/4, \\
\omega_{\mathrm{II}}^A, & \quad\quad\quad J_1/4 \le J_2 \le J_1.
\end{cases}  \label{eq:vds}
\end{align}
The maximum eigenfrequency $\omega_+^A$ always occurs at the quantum number $q = 0$, irrespective of the value of $J_2$. However, the minimum eigenfrequency $\omega_-^A$ can take on two distinct values, either $\omega_{\mathrm{I}}^A$ or $\omega_{\mathrm{II}}^A$, depending upon the crucial coupling ratio $J_2/J_1$. Explicitly, we find
\begin{align}
\label{eq:gsdfdgfhdfgdfgdfgfdgffghfsdfdgf}
\omega_{\mathrm{I}}^A &=  \omega_0 - 2 \left( J_1 - J_2 \right), \\
\omega_{\mathrm{II}}^A &=   \omega_0 - 2 \left( J_2 + \tfrac{J_1^2}{8 J_2} \right). \label{eq:sfdvdffvfe}
\end{align}
The state with the quantum number $q = \pi/d$, residing at the edge of the first Brillouin zone, is associated with the band minimum of $\omega_{\mathrm{I}}^A$ for weaker next-nearest-neighbour couplings satisfying $0 \le J_2 \le J_1/4$, which is demonstrated with the coloured dashed lines in Fig.~\ref{chin}~(a, e) [cf. Eq.~\eqref{eq:gsdfdgfhdfgdfgdfgfdgffghfsdfdgf}]. For stronger next-nearest-neighbour couplings fulfilling $J_1/4 \le J_2 \le J_1$, the band minimum of $\omega_{\mathrm{II}}^A$ instead occurs away from the edge of the Brillouin zone at the specific quantum numbers
\begin{equation}
\label{eq:sdfsdfsdfvdfsdf}
q = \pm \frac{1}{d} \left( \pi - \arctan \left[ \sqrt{ \left( \tfrac{4 J_2}{J_1} \right)^2 -1} \right] \right),
\end{equation}
as marked by the horizontal dashed lines in Fig.~\ref{chin}~(i, m) [cf. Eq.~\eqref{eq:sfdvdffvfe}]. These bandstructural extrema features of Eq.~\eqref{eq:gsdfdgfhfghfsdfdgf} and Eq.~\eqref{eq:vds} [as highlighted by dashed lines throughout the first column of Fig.~\ref{chin}] as well as the additional stationary points [that is, those stationary points not residing at $qd = 0$ or $\pi$, which are represented with coloured circles in panels~(i, m)] together suggest two distinct behavioural regimes for the chain A model, which are separated by the key point in parameter space $J_2 = J_1 /4$.

\begin{figure*}[tb]
 \includegraphics[width=\linewidth]{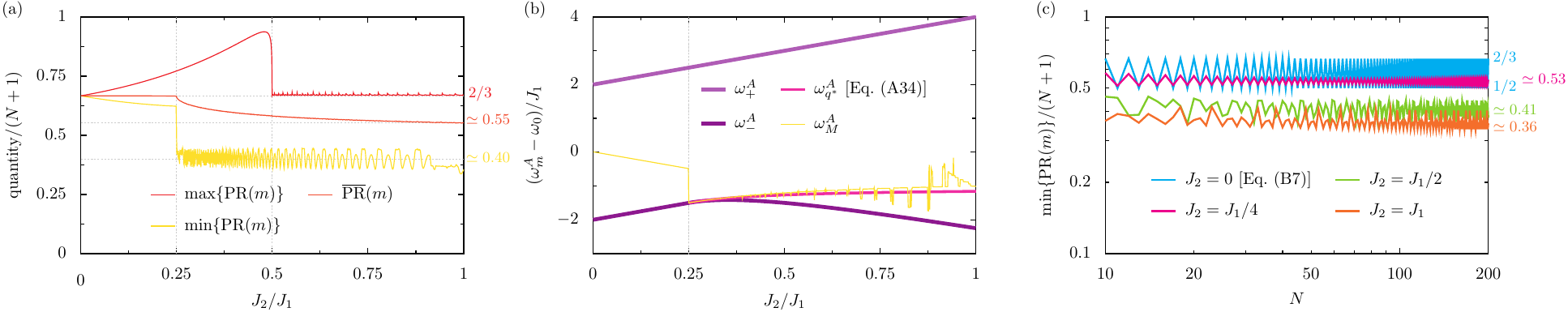}
 \caption{ \textbf{Degrees of localization in the chain A model.} Panel (a): various localization measures, scaled by $N+1$, related to the participation ratio $\mathrm{PR} \left( m \right)$ of a state $m$.  We plot as a function of the next-nearest neighbour coupling $J_2$ (in units of the nearest neighbour coupling $J_1$), and here we consider a finite system of $N = 300$ oscillators. Red line: $\mathrm{max} \{ \mathrm{PR} \left( m \right) \}$, the maximum of the participation ratios of all states $m$. Orange line: $\overline{\mathrm{PR} \left( m \right)}$, the mean value of the participation ratios. Yellow line: $\mathrm{min} \{ \mathrm{PR} \left( m \right) \}$, the minimum of the participation ratios. Panel (b): some chosen eigenfrequencies as a function of $J_2$. Thin yellow line: the specific eigenfrequency $\omega_{M}^A$ corresponding to the state with the index $M$, that with the smallest participation ratio. Pink line: the eigenfrequency $\omega_{q^\ast}^A$, corresponding to the state with the quantum number $q^\ast$, which is associated with the appearance of an additional root in the inverse effective mass in the continuum limit [cf. Eq.~\eqref{eq:gdfrgd}]. Purple lines: $\omega_{\pm}^A$, the extrema of the continuum model band [cf. Eq.~\eqref{eq:gsdfdgfhfghfsdfdgf} and Eq.~\eqref{eq:vds}]. Panel (c): a log-log plot of the minimum of the scaled participation ratio $\mathrm{min} \{ \mathrm{PR} \left( m \right) \} / (N+1)$ as a function of the system size $N$. We consider the coupling ratios $J_2 / J_1 = \{ 0, 1/4, 1/2, 1 \}$ with the cyan, pink, green and orange lines respectively. The result for $J_2 = 0$ (cyan line) is analytic [cf. Eq.~\eqref{eq:grgbrd}]. }
 \label{diss1}
\end{figure*}

Within standard band theory, the group velocity $v_q^A = \partial_q \omega_{q}^A$, a measure of how quickly an excitation of energy $\omega_{q}^A$ moves within the lattice, and the inverse effective mass $1/m_q^A = \partial_q^2 \omega_{q}^A$, a quantifier of how easy it is to accelerate an excitation of energy $\omega_{q}^A$, follow directly from taking derivatives of the dispersion of Eq.~\eqref{eq:gdfdgf2} with respect to the quantum number $q$, like so
\begin{align}
\label{eq:sdf}
v_q^A &= - 2 d \left[ J_1 \sin \left( qd \right) +2 J_2  \sin \left( 2 qd \right) \right], \\
1/m_q^A &= - 2 d^2 \left[ J_1 \cos \left( qd \right) + 4 J_2  \cos \left( 2 qd \right) \right]. \label{eq:sdf2}
\end{align}
These two fundamental quantities are plotted in the second and third columns of Fig.~\ref{chin} respectively, and they correspond to the bandstructures $\omega_{q}^A$ plotted in the adjacent panels of the first column of Fig.~\ref{chin}. In particular, the critical values of $q$ corresponding to additional stationary points in $\omega_{q}^A$ are similarly marked by coloured circles in the panels (j) and (n), which are exactly the additional zeroes in the group velocity $v_q^A$ due to the non-negligible next-nearest coupling $J_2$. Notably, these additional stationary points are also marked by coloured circles in the inverse effective mass $1/m_q^A$ plots of panels (k) and (o), which implies that these states have a rather small effective mass $m_q^A$ and hence that they are more prone to moving around the lattice. Importantly, while the inverse effective mass $1/m_q^A$ is a sign-changing quantity as a function of the quantum number $q$ throughout the third column in Fig.~\ref{chin}, the lowest two panels (k) and (o) have additional roots in the inverse effective mass at certain quantum numbers $q = q^\ast$, which are relatively near to the Brillouin zone edges (that is, they are not the roots in the vicinity of $q d = \pm \pi/2$). These $q^\ast$ modes indeed satisfy $1/m_{q^\ast}^A = 0$, where
\begin{equation}
\label{eq:xcvxcvc}
q^\ast = \pm \frac{1}{d} \left( \pi - \arctan \left[ \frac{ \sqrt{ 128 J_2^2 - 2 J_1 \left( \sqrt{J_1^2 + 128 J_2^2}  + J_1 \right) } }{ \sqrt{J_1^2 + 128 J_2^2} + J_1 } \right] \right).
\end{equation}
These special states with the quantum number $q^\ast$ correspond to excitations with infinite effective masses, which perhaps implies a greater tendency for them to localize. This is discussed quantitatively later on around Eq.~\eqref{eq:gdfrgd} and in Fig.~\ref{diss1}~(b) when a finite array is considered, and the underpinning mechanism is linked to the ideas of Sajeev John and co-workers~\cite{John1991}.

The density of states $g \left( \omega \right)$ counts the number of states at a given frequency $\omega$. It is defined for a system of $N$ energy levels, each residing at a certain frequency $\omega_n$, via the informal summation formula
\begin{equation}
\label{eq:gsdfdfsdfdgf}
g \left( \omega \right) = \sum_{n=1}^N \delta \left( \omega - \omega_n \right),
\end{equation}
where $\delta (x)$ is Dirac's delta function, and where the integral over the density of states $\int_{-\infty}^\infty g \left( \omega \right) \mathrm{d}\omega = N$ indeed recovers the total number of energy levels. With regard to the considered Hamiltonian operator $\hat{H}_A$ of Eq.~\eqref{eq:gdfdgf}, where the quantum number $q = 2 \pi m / N d$ due to the employed periodic boundary conditions, one may rewrite Eq.~\eqref{eq:gsdfdfsdfdgf} in the continuum limit of $N \to \infty$ oscillators as the density of states integral
\begin{equation}
\label{eq:gsdfdsdfwwwdfsdfdgf}
g^A \left( \omega \right) = \frac{N d}{2 \pi} \int_{-\frac{\pi}{d}}^{\frac{\pi}{d}} \delta \left( \omega - \omega_{q}^A \right) \mathrm{d}q,
\end{equation}
where the integral is over the whole of the first Brillouin zone, and where the eigenfrequencies $\omega_{q}^A$ are provided by the dispersion relation of Eq.~\eqref{eq:gdfdgf2}. The Dirac delta function of some function $f(x)$ is conveniently expressable using the identity
\begin{equation}
\label{eq:gsdfdfgfdfsdfdgf}
\delta \left[ f \left( x \right) \right] = \sum_n \frac{\delta \left( x - x_n \right)}{ | f' \left( x_n \right) | },
\end{equation}
where $x_n$ are the roots of the function $f(x)$ so that $f(x_n) = 0$, and where the first derivative $f'(x) = \frac{\mathrm{d}}{\mathrm{d}x} f(x)$. Working in the dimensionless variable $x$, and introducing the dimensionless quantity $a$ and the dimensionless coupling strength parameter $b$, as 
\begin{align}
\label{eq:sdfsdfdfsvfs}
x &= q d, \\
a &= \frac{\omega - \omega_0}{2 J_1},  \label{eq:aadcs} \\
b &= \frac{J_2}{J_1},  \label{eq:aadcs2}
\end{align}
we may rewrite the argument of the Dirac delta function appearing in the density of states of Eq.~\eqref{eq:gsdfdsdfwwwdfsdfdgf} as follows
\begin{equation}
\label{eq:sdfvfs}
f \left( x \right) = \omega - \left[ \omega_0 + 2 J_1 \cos \left( x \right) + 2 J_2 \cos \left( 2 x \right) \right]. 
\end{equation}
Meanwhile the derivative of this function, and its root equation appearing within the delta function identity of Eq.~\eqref{eq:gsdfdfgfdfsdfdgf}, can be compactly expressed in the forms
\begin{align}
f' \left( x \right) &= 2 J_1 \left[ \sin \left( x \right) + 2 b \sin \left( 2 x \right) \right], \label{eq:sdfvfs2} \\
a &= \cos \left( x_n \right) +  b \cos \left( 2 x_n \right). \label{eq:sdfvfs3}
\end{align}
Upon switching to the new variable $z_n = \mathrm{e}^{\mathrm{i} x_n}$, the trigonometric Eq.~\eqref{eq:sdfvfs3} is equivalent to the following quartic equation
\begin{equation}
\label{eq:gsdfdsfsdfsdfssdfwwwdfsdfdgf}
b z_n^4 + z_n^3 - 2 a z_n^2 + z_n + b = 0.
\end{equation}
The four solutions $z_{n}$ of the quartic Eq.~\eqref{eq:gsdfdsfsdfsdfssdfwwwdfsdfdgf} are fortunately rather compact, being
\begin{align}
\label{eq:ssdfsdfdfvfs}
z_{1, 2} &= \frac{-1 -\sqrt{1+ 8 b \left( b + a \right)} \pm \sqrt{2 - 8 b \left(  b - a \right) + 2 \sqrt{ 1 + 8 b \left( b + a \right) } } }{4 b}, \\
z_{3, 4} &= \frac{-1 + \sqrt{1+ 8 b \left( b + a \right)} \pm \sqrt{2 - 8 b \left( b - a \right) - 2 \sqrt{ 1 + 8 b \left( b + a \right) } } }{4 b}. \label{eq:ssdfsdfdfvfs2}
\end{align}
The solutions of Eq.~\eqref{eq:ssdfsdfdfvfs} and Eq.~\eqref{eq:ssdfsdfdfvfs2} imply that there are only two real roots $x_n$ in the case of weak next-nearest coupling $0 \le J_2 \le J_1/4$, so that the integral of Eq.~\eqref{eq:gsdfdsdfwwwdfsdfdgf} can be readily carried out. This procedure leads to the density of states
\begin{equation}
\label{eq:sfvfdfsdffsfdsgrtr}
g^A \left( \omega \right) = \frac{N}{2 \pi J_1} F_+ \left( a, b \right) \Theta \left( \omega - \omega_{\mathrm{I}}^A \right) \Theta \left( \omega_+^A - \omega \right), 
\end{equation}
which indeed satisfies $\int_{-\infty}^\infty g^A \left( \omega \right) \mathrm{d}\omega = N$. Within Eq.~\eqref{eq:sfvfdfsdffsfdsgrtr}, $\Theta (x)$ is the Heaviside step function, which takes into account the band extrema at $\omega_{+}^A$ and $\omega_{\mathrm{I}}^A$ in this weak coupling regime [cf. Eq.~\eqref{eq:gsdfdgfhfghfsdfdgf} and Eq.~\eqref{eq:gsdfdgfhdfgdfgdfgfdgffghfsdfdgf}], while the two parameters $a$ and $b$ are defined in Eq.~\eqref{eq:aadcs} and Eq.~\eqref{eq:aadcs2} respectively. In writing down the density of states $g^A \left( \omega \right)$, we have introduced the auxiliary function $F_+ \left( a, b \right)$, which is defined through the expression for $F_\pm$ as
\begin{equation}
\label{eq:sfvfdfdgfgdsdffsfdsgrtr}
F_\pm \left( a, b \right) = \frac{4 b}{ \sqrt{ 2 + 16 b \left( a + b\right) } \sqrt{ 4 b \left( b - a \right) - 1 \pm \sqrt{ 1 + 8 b \left( a + b \right) } } }. 
\end{equation}
Notably, the auxiliary function $F_+ \left( a, b \right)$ collapses into much simpler forms for the following special values of $b$, which are the limiting cases
\begin{align}
\label{eq:sdasdsdsdsfafvfs}
F_+ \left( a, 0 \right) &= \frac{1}{\sqrt{1-a^2}}, \\
F_+ \left( a, \tfrac{1}{4} \right) &= \frac{2}{\sqrt{3 + 4 a} \sqrt{2 \sqrt{6 + 8 a} - 4 a - 3} }. \label{eq:dfsvfdfh6u6}
\end{align}
The result of Eq.~\eqref{eq:sdasdsdsdsfafvfs} allows for the reproduction of the famous nearest-neighbour coupling only ($J_2 = 0$) result, which features the celebrated van Hove inverse square root singularities at the band edges. The expression of Eq.~\eqref{eq:dfsvfdfh6u6} is reached at the largest value of the weak next-nearest coupling regime, $J_2 = J_1/4$. The case of strong next-nearest coupling $J_1/4 \le J_2 \le J_1$ proceeds in a similar fashion to Eq.~\eqref{eq:sfvfdfsdffsfdsgrtr}, but now all four solutions of Eq.~\eqref{eq:ssdfsdfdfvfs} and Eq.~\eqref{eq:ssdfsdfdfvfs2} need to be employed thanks to the additional stationary points. The analogous calculation leads directly to the density of states 
\begin{equation}
\label{eq:sfvfdfnlsdsdfsdfffsfdsgrtr}
g^A \left( \omega \right) = \frac{N}{2 \pi J_1} \left[ F_+ \left( a, b \right) \Theta \left( \omega - \omega_{\mathrm{II}}^A \right) \Theta \left( \omega_+^A - \omega \right)
+ F_- \left( a, b \right) \Theta \left( \omega - \omega_{\mathrm{II}}^A \right) \Theta \left(  \omega_{\mathrm{I}}^A - \omega \right) \right],
\end{equation}
where the auxiliary functions $F_{\pm} \left( a, b \right)$ are defined in Eq.~\eqref{eq:sfvfdfdgfgdsdffsfdsgrtr}. Notably, the effect of the additional stationary points arising due to significant next-nearest coupling $J_2$ has led to the appearance of a second term in Eq.~\eqref{eq:sfvfdfnlsdsdfsdfffsfdsgrtr} as compared to Eq.~\eqref{eq:sfvfdfsdffsfdsgrtr}, which creates additional van Hove divergencies in the overall density of states. Using the results of Eq.~\eqref{eq:sfvfdfsdffsfdsgrtr} and Eq.~\eqref{eq:sfvfdfnlsdsdfsdfffsfdsgrtr}, we plot the density of states $g^A ( \omega )$ in the fourth column of Fig.~\ref{chin}. The transition from two [panels (d) and (h)] to three [panels (l) and (p)] non-smooth points is readily seen with increasing $J_2$ as the column descends.

The band theory results discussed here for the chain A model reduce in the limit of nearest-neighbour coupling only ($J_2 \to 0$) to well-known results. In particular, the bandstructure and its extrema, the group velocity, inverse effective mass and the density of states are described by the pleasingly short expressions
\begin{align}
\label{eq:sdsdfsdffsdfs}
\omega_{q}^A &= \omega_0 + 2 J_1 \cos \left( qd \right),  \\
\omega_+^A &= \omega_0 + 2 J_1, \\
 \omega_-^A &= \omega_0 - 2 J_1, \\
v_q^A &=  - 2 J_1 d \sin \left( qd \right), \label{eq:gdsf} \\
1/m_q^A &=  -2 J_1 d^2 \cos \left( qd \right), \\
g^A \left( \omega \right) &= \frac{N}{2 \pi J_1} \frac{\Theta \left( \omega - \omega_-^A \right) \Theta \left( \omega_+^A - \omega \right)}{\sqrt{1- \left( \frac{\omega - \omega_0 }{2 J_1} \right)^2 }},  \label{eq:fsdvsdghhhy}
\end{align}
which all appear graphically in the top row of Fig.~\ref{chin} [cyan lines in panels (a, b, c, d)], where we consider $J_2 = 0$ as the baseline case from which the more complicated behaviours in the lower rows eventually emerge. In particular, the cosine band of Eq.~\eqref{eq:sdsdfsdffsdfs} leads to standard Van Hove singularities in the density of states of Eq.~\eqref{eq:fsdvsdghhhy} and Fig.~\ref{chin}~(d), which are typical in one-dimensional systems. Clearly from Eq.~\eqref{eq:gdsf}, there are only stationary points in the bandstructure at $qd = 0$ and $qd = \pi$, while zeroes of the inverse effective mass only occur at $qd = \pm \pi/2$, as shown graphically in Fig.~\ref{chin}~(b, c). In a future study, it may be interesting to consider the influence of the counter-rotating terms~\cite{Lamata2019, Toghill2022} discarded from the coupling part of the Hamiltonian, as well as to examine the impact of including an interaction term~\cite{Zurita2019, Azcona2021} in the Hamiltonian in order to probe a nonlinear array of anharmonic oscillators.

\begin{figure*}[tb]
 \includegraphics[width=\linewidth]{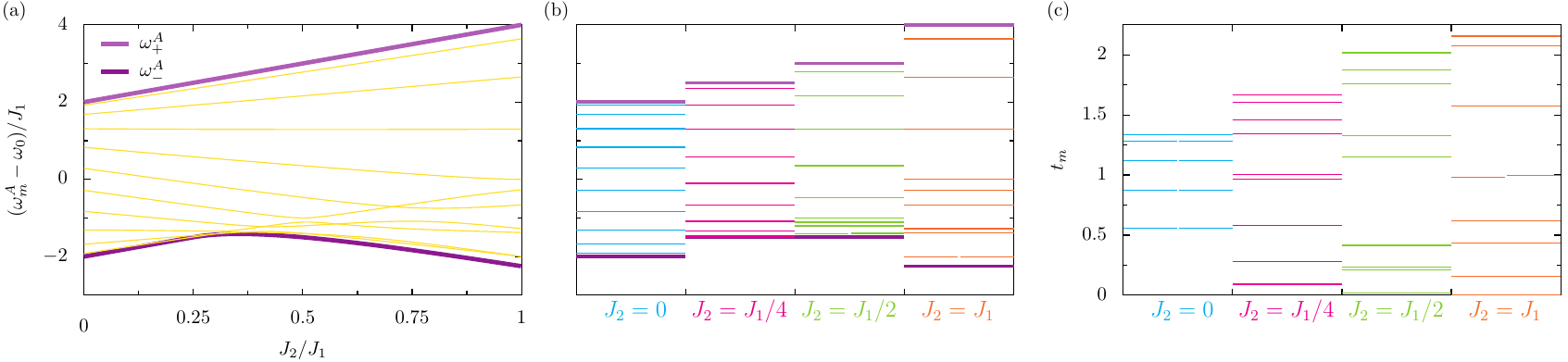}
 \caption{ \textbf{Eigenfrequencies of the chain A model for a finite system with open boundary conditions.} Panel (a): the eigenfrequencies $\omega_{m}^A$ for a finite system of $N = 10$ oscillators, as a function of the next-nearest neighbour coupling $J_2$ (in units of the nearest neighbour coupling $J_1$), as marked by the yellow lines. Thick purples lines: $\omega_{\pm}^A$, the extrema of the equivalent band formed in the continuum limit $N \to \infty$ [cf. Eq.~\eqref{eq:gsdfdgfhfghfsdfdgf} and Eq.~\eqref{eq:vds}]. Panel (b): the specific eigenfrequencies $\omega_{m}^A$, taken from panel (a) at the chosen values of $J_2 / J_1 = \{ 0, 1/4, 1/2, 1 \}$, as associated with the cyan, pink, green and orange lines respectively. Purple lines: the extrema $\omega_{\pm}^A$ in the continuum limit. Panel (c): the relative energy level spacings $t_m$ between the eigenfrequencies $\omega_{m}^A$ taken from panel (b).}
 \label{levels}
\end{figure*}

Thus far, we have exploited periodic boundary conditions and considered the limit of an infinitely large system ($N \to \infty$) in order to study some fundamental band properties of chain A analytically. We shall now consider open boundary conditions with the Hamiltonian operator $\hat{H}_A$ for a finite system of $N$ oscillators, which allows us to study the degrees of localization and energy level statistics of the model. The theoretical treatment requires the diagonalization of the $N \times N$ matrix form of Eq.~\eqref{eq:sdfsdfsfd}, which results in a finite number of eigenfrequencies $\omega_{m}^A$ for the countable number of states, which are labelled from $m = \{ 1, 2, ..., N \}$. The values of the associated eigenvectors, on oscillator $n$ and for a state $m$, are given by the amplitude $c_n^A (m)$. We show some typical eigenfrequency results, as a function of the coupling ratio $J_2/J_1$, for a chain of $N = 10$ oscillators with the yellow lines in Fig.~\ref{levels}~(a). As a guide for the eye, we include the continuum band edge results of $\omega_{\pm}^A$ as the thick purple lines, using Eq.~\eqref{eq:gsdfdgfhfghfsdfdgf} and Eq.~\eqref{eq:vds}, which act as the energetic bounds. We see from Fig.~\ref{levels}~(a) how the collection of ten eigenfrequencies $\omega_{m}^A$ evolve with increasing $J_2/J_1$, and in particular we notice the crucial point around $J_2 = J_1/4$, where the minimum eigenfrequency finally starts to lower due to the appearance of a new stationary point in the analogous infinite system [cf. Eq.~\eqref{eq:vds}], and where several energy levels start to cluster together, which is important for what follows.

We quantify the degrees of localization of the states $m$ sustained by chain A using the participation ratio $\mathrm{PR} \left( m \right)$, as defined in Eq.~\eqref{eq:sdfddwsdfs}. Typically, this measure takes the form of $\mathrm{PR} \left( m \right) = (2/3)(N+1)$ or similar in conventional, one-dimensional tight-binding systems [as is shown later on in Eq.~\eqref{eq:sdfddwcasasssdfs}]. The linear dependence on $N$ is characteristic of all extended states, while the prevalent prefactor of $2/3$ arises from the geometry of the linear chain with open boundary conditions. In Fig.~\ref{diss1}~(a) we are interested in the three measures, all of which are linked to the participation ratio and are scaled by the convenient factor $N+1$, as a function of the coupling ratio $J_2/J_1$. The red line in Fig.~\ref{diss1}~(a) denotes $\mathrm{max} \{ \mathrm{PR} \left( m \right) \}$, the maximum of the participation ratios (found after consideration of all states $m$). With vanishing $J_2$ and for $J_2 > J_1/2$, this quantity is essentially the constant $\mathrm{max} \{ \mathrm{PR} \left( m \right) \} / (N+1) \simeq 2 / 3$ [this relation is exact for $J_2 = 0$]. This implies a linear relationship between the participation ratio and the system size, while the $2 / 3$ prefactor is typical of extended states. Interestingly, for $0 < J_2 < J_1/2$ a larger prefactor associated with $\mathrm{max} \{ \mathrm{PR} \left( m \right) \}$ is demonstrated, which perhaps indicates the presence of some unusually super-extended states due to the non-negligible next-nearest neighbour hoppings. Similarly, the yellow line in Fig.~\ref{diss1}~(a) denotes $\mathrm{min} \{ \mathrm{PR} \left( m \right) \}$, graphing the minimum of the participation ratio (over all states $m$) scaled by $N+1$. The additional stationary points suddenly occurring around $J_2 = J_1/4$ are seemingly correlated with an unusually low prefactor of $\sim 0.4$ in this case, which may perhaps be termed sub-extended states since the linear scaling with $N$ is maintained, but less than one half of the total number of sites are felt by the state. Finally, the mean of the participation ratio $\overline{\mathrm{PR} \left( m \right)}$, calculated as the average over all $N$ states, is represented by the orange line in Fig.~\ref{diss1}~(a). The orange curve suggests that the average state does not behave all that differently below and above the critical coupling of $J_2/J_1 = 1/4$.

The calculations leading to the results of Fig.~\ref{diss1}~(a) were carried out for a finite chain of $N = 300$ oscillators, which also led to the results of panel Fig.~\ref{diss1}~(b). There we plot some key eigenfrequencies as a function of the coupling ratio $J_2 / J_1$. Similarly to Fig.~\ref{levels}~(a), the purple lines mark the continuum band edges $\omega_{\pm}^A$ as a guide for the eye at the energetic bounds. More importantly, the yellow line in Fig.~\ref{diss1}~(b) tracks the special eigenfrequency $\omega_{M}^A$, coming from the particular state $m=M$ with the smallest participation ratio $\mathrm{min} \{ \mathrm{PR} \left( m \right) \}$. For the important coupling regime of $J_1 /4 < J_2 < J_1$, where the participation ratio of the state $M$ is appreciably smaller than for all other states [cf. Fig.~\ref{diss1}~(a)], the key eigenfrequency $\omega_{M}^A$ is remarkably seen to be not too dissimilar to the analytic expression
\begin{equation}
\label{eq:gdfrgd}
\omega_{q^\ast}^A = \omega_0 - \frac{3 J_1}{32 J_2} \left( J_1 +  \sqrt{J_1^2 + 128 J_2^2} \right), 
\end{equation}
as plotted with the pink line in Fig.~\ref{diss1}~(b). We arrived at Eq.~\eqref{eq:gdfrgd} by substituting the specific the quantum number $q = q^\ast$ [cf. Eq.~\eqref{eq:xcvxcvc}], which is associated with a zero in the inverse effective mass $1/m_q^A$ from Eq.~\eqref{eq:sdf2}, into the bandstructure of Eq.~\eqref{eq:gdfdgf2}. The approximate likeness of $\omega_{M}^A$ and Eq.~\eqref{eq:gdfrgd}, as suggested by Fig.~\ref{diss1}~(b), can be interpreted as this particular excitation with $q^\ast$ having a large effective inertia, such that the state is much more resistant to extending over the entire one-dimensional array of coupled oscillators.

We confirm the extended nature of all of the states sustained in the chain A model in Fig.~\ref{diss1}~(c), which plots the minimum of the scaled participation ratio $\mathrm{min} \{ \mathrm{PR} \left( m \right) \} / (N+1)$ as a function of the overall system size $N$. We consider four values of $J_2$, corresponding to the four cases considered previously in Fig.~\ref{chin}. Clearly, all of the cases quickly tend towards constant values, revealing the linear dependence of this localization measure with the total number of oscillators $N$ in the chain. In particular, the case of $J_2 = 0$ (cyan line) is describable analytically [as shown later on in Eq.~\eqref{eq:grgbrd}, prompting the choice of scaling with $N+1$ rather than simply $N$] and it exhibits a triangular waveform due to an even-odd relationship with the system size $N$.

\begin{figure*}[tb]
 \includegraphics[width=\linewidth]{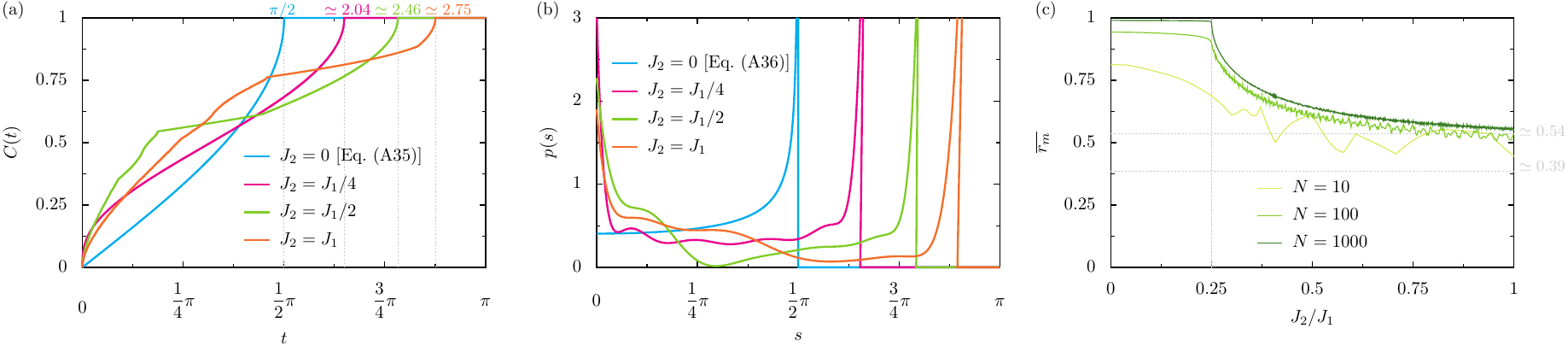}
 \caption{  \textbf{Energy level statistics of the chain A model.} Panel (a): the cumulative distribution function  $C \left( t \right)$ for the adjacent energy level spacings, for the four coupling ratios $J_2 / J_1$ also considered in Fig.~\ref{diss1}~(c). Panel (b): the probability density function $p \left( s \right)$ associated with the results of panel (a), so that $\int_0^t p \left( s \right) \mathrm{d} s = C \left( t \right)$. We consider a finite system of $N = 15000$ oscillators in panels (a) and (b) for the cases with $J_2 \ne 0$, otherwise we use the analytic expressions of Eq.~\eqref{eq:bomdaws} and Eq.~\eqref{eq:bomdaws2} for the case of $J_2 = 0$ [cyan lines in both panels]. Panel (c): the mean value of the ratio of consecutive energy level spacings $\overline{r_m}$, as a function of $J_2$. We consider three increasingly large finite systems with $N = \{ 10, 100, 1000\}$ oscillators. Dashed grey lines: the values $2 \ln 2 -1 \simeq 0.39$ (associated with Poissonian level statistics and a localized phase) and $4 - 2 \sqrt{3} \simeq 0.54$ (associated with Gaussian orthogonal ensembles and an extended phase). }
 \label{diss2}
\end{figure*}

The energy level statistics of chain A are investigated in Fig.~\ref{diss2}. In particular, we consider two closely related quantities describing energy level correlations in Fig.~\ref{diss2}~(a) and (b)~\cite{Mehta2014, Guhr1998}. Meanwhile Fig.~\ref{diss2}~(c) is focussed on the statistics of consecutive energy level spacings~\cite{Oganesyan2007, Atas2013}, as was also investigated in Fig.~\ref{numer}~(c) for the two-leg ladder model of the main text. We previously described with yellow lines in Fig.~\ref{levels}~(a) the eigenfrequencies $\omega_{m}^A$, as a function of the coupling ratio $J_2/J_1$, for a relatively small chain of $N = 10$ oscillators. Taking four cuts through Fig.~\ref{levels}~(a) at the coupling ratios $J_2/J_1 = \{ 0, 1/4, 1/2, 1\}$ results in the energy level diagram sketched in Fig.~\ref{levels}~(b), where some level clusterings become more obviously apparent than in Fig.~\ref{levels}~(a). With these energy levels $E_m = \omega_{m}^A$, ordered such that $m = 1$ corresponds to the smallest eigenfrequency and ascending to the largest eigenfrequencies when $m = N$, we may define the $N-1$ energy level separations $S_m = E_{m+1} - E_{m}$, where the index $m = \{ 1, 2,  ..., N-1 \}$. The mean energy level spacing is the average of these energetic distances $D = \overline{S_m}$, such that one may define the $N-1$ normalized energy level spacings $t_m = S_m/D$. We plot these normalized and dimensionless energy level spacings $t_m$ in Fig.~\ref{levels}~(c), corresponding to the eigenfrequencies of Fig.~\ref{levels}~(b). In the thermodynamic limit of $N \to \infty$, the distribution of $t_m$ can then be described within the statistical theory of energy levels~\cite{Mehta2014, Guhr1998} as we now discuss.

The cumulative distribution function $C(t)$ of the adjacent energy levels, as a function of the continuum energy level spacing $t$, is plotted in Fig.~\ref{diss2}~(a) for four representative values of $J_2$ within the chain A model. There are no energy level spacings with a spacing below a vanishing amount such that $C(t \to 0) = 0$, while for some sufficiently large value of the dimensionless energy level spacing $t = T$ all energy level spacings in the chain A model should be smaller than it (due to the collection of energy levels packing themselves into an effective band), such that $C(t \ge T) = 1$. The simplest case of $J_2 = 0$ is denoted by the cyan line in Fig.~\ref{diss2}~(a), whose cumulative distribution curve is obtained analytically with the exact inverse trigonometric expression
\begin{equation}
\label{eq:bomdaws} 
C \left( t \right) = \frac{2}{\pi} \arcsin \left( \frac{2 t}{\pi} \right) \Theta \left( \tfrac{\pi}{2} - t \right) + \Theta \left( t - \tfrac{\pi}{2} \right),  
\end{equation}
where the threshold dimensionless energy level spacing $T = \pi/2$, as is marked at the top of Fig.~\ref{diss2}~(a). The derivation of Eq.~\eqref{eq:bomdaws} is provided later on in App.~\ref{Sec:chainB}, where the chain B model is analyzed. The results for three other cases with $J_2 \ne 0$, corresponding to the values considered in Fig.~\ref{levels}~(b, c), are shown in Fig.~\ref{diss2}~(a) after a numerical calculation for a finite system composed of N = 15000 oscillators. This leads to successively larger values energy level spacings $T$ for which $C(T) = 1$ first occurs, due to the increasingly large bandwidths of the effective bands formed by the collections of energy levels. These approximate threshold $T$ values are noted at the top of Fig.~\ref{diss2}~(a).

The probability density functions $p(s)$ associated with the four curves of Fig.~\ref{diss2}~(a), such that the integral $\int_0^t p \left( s \right) \mathrm{d} s = C \left( t \right)$, are plotted in Fig.~\ref{diss2}~(b). The probability density function $p(s)$ in the simplest case of $J_2 = 0$ follows from taking the derivative of Eq.~\eqref{eq:bomdaws} with respect to $t$ and evaluating the result at the dimensionless  value $s$, leading to the neat formula 
\begin{equation}
 \label{eq:bomdaws2}
p \left( s \right) = \frac{2}{\pi} \frac{ \Theta \left( \frac{\pi}{2} - s \right) }{\sqrt{ \left( \frac{\pi}{2} \right)^2  - s^2}},
\end{equation}
which indeed satisfies the normalization $\int_0^\infty p \left( s \right) \mathrm{d} s = 1$ in order to conserve the total probability. Furthermore, the first moment obeys $\int_0^\infty s p \left( s \right) \mathrm{d} s = 1$, since the mean level spacing was normalized to unity. The inverse-square root distribution of Eq.~\eqref{eq:bomdaws2} is shown by the cyan line Fig.~\ref{diss2}~(b), demonstrating a lack of energy level repulsion since $p(s \to 0) = (2/\pi)^2 \simeq 0.405$. This is followed by an increasing likelihood of larger level separations until the critical value of $s = \pi/2$ is reached, after which there is a step function drop-off to zero due to the finite bandwidth of the chain A model. The rough numerical results for the three nonzero values of $J_2$ considered in Fig.~\ref{diss2}~(a) are also shown in Fig.~\ref{diss2}~(b), which display significantly enhanced clusterings of the energy levels due to the $s \to 0$ behaviour, as was already hinted at in the level spacing diagram of Fig.~\ref{levels}~(c). The critical behaviour (abrupt dropping to zero) of the $J_2 \ne 0$ cases in Fig.~\ref{diss2}~(b) appears at increasingly large values of $s$, as is expected from the analytic case of $J_2 = 0$ and Eq.~\eqref{eq:bomdaws2}, where the criticality lies at $s = \pi/2$.

For a given spectrum, the ratio of consecutive energy level spacings $r_m$, as defined by Eq.~\eqref{eq:dfsf}, provides a measure of the correlations between adjacent energy level gaps~\cite{Oganesyan2007}. The mean value $\overline{r_m}$ provides a certain sense of the degree of chaoticity, being essentially unity in ordered models and dropping to $\overline{r_m} = 2 \ln 2 -1 \simeq 0.39$ for models with Poissonian level statistics, which sometimes implies a more localized phase~\cite{Atas2013, Torres2019}. The evolution of mean ratio $\overline{r_m}$ for the chain A model with the coupling ratio $J_2/J_1$ is depicted in Fig.~\ref{diss2}~(c). We consider the results for three increasingly large magnitudes of the finite system size $N$ with increasingly dark green lines. Strikingly, while for weak next-nearest couplings $J_2 < J_1 / 4$ the thermodynamic limit of $N \to \infty$ sees the mean ratio approach $\overline{r_m} = 1$, corresponding to an ordered system, the regime of strong next-nearest couplings $J_1/4 < J_2 \le J_1$ is characterized by a decay from unity. This drop-off starts abruptly at the key coupling $J_2 = J_1 / 4$, falling towards a value of $\overline{r_m} \simeq 0.55$ when $J_2 = J_1$ (in the thermodynamic $N \to \infty$ limit). This mean value $\overline{r_m}$ is somewhat similar to the case of a Gaussian orthogonal ensemble, where $\overline{r_m} = 4 - 2 \sqrt{3} \simeq 0.54$~\cite{Tarquini2017}, suggesting a kind of extended phase in the chain A model which is consistent with the participation ratio results of Fig.~\ref{diss1}.
\\


\renewcommand{\theequation}{B \arabic{equation}}
\section{The chain B model}
\label{Sec:chainB}


In this appendix we consider the chain B model as defined by the Hamiltonian operator $\hat{H}_B$ of Eq.~\eqref{eq:sdfsdssfdsfsfd}, describing a finite linear chain of $N$ quantum harmonic oscillators (each of resonance frequency $\omega_0$) interacting via nearest-neighbour coupling (of strength $J_1$) only [see the sketch of chain B in Fig.~\ref{wfs}~(a)]. In particular, we are interested in analytical descriptions of the various degrees of localization arising in this simple system, as judged by the participation ratio~\cite{Torres2019, Schiulaz2018}, and the energy level statistics resulting from the chain B Hamiltonian, as governed by the probability density function for its adjacent energy level spacings~\cite{Zimmermann1987, Pandey1991, Chakrabarti2003}. The mathematical results provided here provide an understanding of the more general chain A model [cf. App.~\ref{Sec:chainA}] in the limit of vanishing $J_2$, and also impart some intuition about the behaviour of the more complicated chain A-B model considered throughout the main text.

Diagonalization of the finite system described by Eq.~\eqref{eq:sdfsdssfdsfsfd}, with the implied open boundary conditions, is best carried out by employing the following discrete sine transform (and its inverse) for the bosonic annihilation and creation operators $b_n$ and $b^\dagger_n$, where
\begin{align}
\label{eq:hfnvn}
b_n &= \sqrt{\tfrac{2}{N+1}} \sum_{k_m} \sin \left( n k_m d \right) b_{k_m}, \\
   b_{k_m} &= \sqrt{\tfrac{2}{N+1}} \sum_{n =1}^N \sin \left( n k_m d \right) b_{n}, \label{eq:hfndfvn}
\end{align}
where we have introduced the quantum number $k_m$, where the dimensionless quantity $k_m d = \pi m / (N +1)$ and the integer $m \in [1, N]$. Substitution of Eq.~\eqref{eq:hfnvn} into Eq.~\eqref{eq:sdfsdssfdsfsfd} yields the diagonalized Hamiltonian operator $\hat{H}_B$, and its associated eigenfrequencies $\omega_{k_m}^B$, as follows
\begin{align}
\label{eq:sfdfsdf}
\hat{H}_B &= \sum_{k_m} \omega_{k_m}^B b_{k_m}^\dagger b_{k_m}, \\
   \omega_{k_m}^B &= \omega_0 + 2 J_1 \cos \left( k_m d \right).  \label{eq:sfdfdfsdfsdsdbbbsdf}
\end{align}
Clearly, in the thermodynamic limit of $N \to \infty$ the eigenfrequencies of Eq.~\eqref{eq:sfdfdfsdfsdsdbbbsdf} form a continuous band of allowed energies [cf. Eq.~\eqref{eq:gdfdgf2} with $J_2 = 0$]. The probability amplitudes $c_n^B (m)$ of each state $m$, for every site $n$ in the chain, follow from the diagonalizing operator $b_{k_m}$ of Eq.~\eqref{eq:hfndfvn} as $c_n^B (m) = \bra{1_n} b_{k_m}^\dagger \ket{ \mathrm{vac} }$, where $\ket{ \mathrm{vac} } = \ket{0, \ldots, 0}$ is the vacuum state. We use the notation that $\ket{ 1_n }$ corresponds to a single excitation on site $n$ only, so that $\ket{ 1_1 } = \ket{1, \ldots,  0}$ and $\ket{ 1_N } = \ket{0, \ldots,  1}$ for example. Then the explicit form of the probability amplitudes $c_n^B (m)$ is
\begin{equation}
\label{eq:dsfdfsf}
c_n^B \left( m \right) = \sqrt{\tfrac{2}{N+1}} \sin \left( n k_m d \right).
\end{equation}
Substitution of Eq.~\eqref{eq:dsfdfsf} into the formula of Eq.~\eqref{eq:sdfddwsdfs} yields the exact participation ratio $\mathrm{PR} \left( m \right)$ for a state $m$ as follows
\begin{equation}
\label{eq:sdfddwcasasssdfs}
\mathrm{PR} \left( m \right) =  \begin{cases}
   \frac{2}{3} \left( N + 1 \right),  & \quad \quad m \ne \frac{N+1}{2}, \\
     \frac{1}{2} \left( N + 1 \right), & \quad \quad m = \frac{N+1}{2}, 
  \end{cases}. 
\end{equation}
As expected, all of the states $m$ are extended due to the explicit linear relationship between the participation ratio and the system size $N$. Essentially, the relationship $\mathrm{PR} \left( m \right) \simeq (2/3) N$ holds for almost all states (the sole exception is for the single state with the eigenfrequency exactly at $\omega_0$, which only occurs for a chain with an odd number $N$ of oscillators). This general prefactor of $2/3$ is typical of extended states and its fingerprints also appear in the chain A-B model via the dark green line in Fig.~\ref{tony}~(c) and in the blue region in Fig.~\ref{numer}~(a), as well as for the chain A model via the red line (for $J_2 \ge J_1/2$) in Fig.~\ref{diss1}~(a). Upon dividing the result of Eq.~\eqref{eq:sdfddwcasasssdfs} by the convenient number $N+1$, it follows that the minimum (taken over all all states $m$) possible scaled participation ratio for this chain B system is simply
\begin{equation}
\label{eq:grgbrd}
\frac{ \mathrm{min} \{ \mathrm{PR} \left( m \right) \}}{N+1} =  \begin{cases}
  1/2, & \quad  N~\text{odd}, \\
   2/3,  & \quad  N~\text{even}, 
  \end{cases}. 
\end{equation}
The formula of Eq.~\eqref{eq:grgbrd} defines a triangular waveform relationship of the least extended state with increasing system size $N$, as was already demonstrated graphically by the cyan line in Fig.~\ref{diss1}~(c).

\begin{figure*}[tb]
 \includegraphics[width=\linewidth]{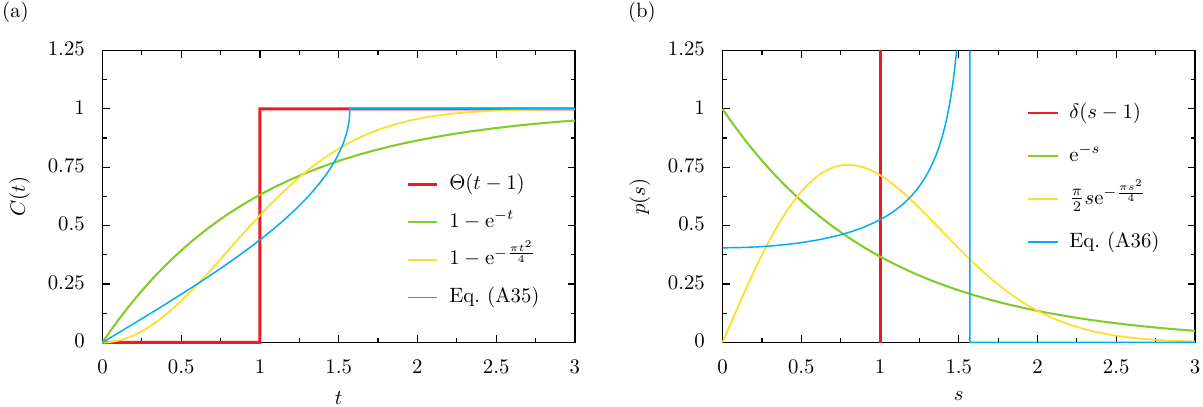}
 \caption{ \textbf{A comparison of the energy level statistics of the chain B model with some celebrated models.} Panel (a): the cumulative distribution function $C (t)$ as a function of the dimensionless energy level spacing $t$ for four different models. Panel (b): the probability density functions $p (s)$ corresponding to the curves of panel (a). Red lines: the picket fence model, where the equal energy level spacing leads to a Dirac delta function distribution. Green lines: the uncorrelated energy levels model, where the random energy level spacing leads to a Poisson distribution. Yellow lines: a chaotic model, where the Gaussian orthogonal ensemble leads to the simplest type of Wigner distribution. Cyan lines: the chain B model, as described analytically by the cumulative and density functions of Eq.~\eqref{eq:bomdaws} and Eq.~\eqref{eq:bomdaws2} respectively.}
 \label{wig}
\end{figure*}

Let us now consider some statistical properties of the energy levels~\cite{Mehta2014, Guhr1998} of the chain B model, using the exact eigenfrequencies $\omega_{k_m}^B$ from Eq.~\eqref{eq:sfdfdfsdfsdsdbbbsdf}. We start by reordering these $N$ energy levels, now relabelled as $E_{m}$ for simplicity, in the ascending order $E_1 < E_2 < E_3 ...$ like so
\begin{equation}
\label{eq:ssdfdfsfdsfdc}
  E_{m} = \omega_0 + 2 J_1 \cos \left(\pi  \frac{  N + 1 - m  }{N+1} \right),
\end{equation}
where the index $m = \{ 1, 2, ..., N\}$. There are then $N-1$ adjacent energy level separations $S_m = E_{m+1} - E_{m}$, which read
\begin{equation}
\label{eq:ssdfdfsfdsfxvcvxdc}
   S_{m} = 2 J_1 \biggl\{ \cos \left( \pi \frac{N - m}{N+1} \right) - \cos \left( \pi \frac{N+1 - m}{N+1} \right) \biggl\},
\end{equation}
whose index spans $m = \{ 1, 2, ..., N-1\}$. The mean value of these adjacent energy level separations $D = \overline{S_m}$ is given by
\begin{equation}
\label{eq:ssdfdfsfdsfxsdasdavcvxdc}
   D = \frac{2 J_1}{N-1} \biggl\{ \cos \left( \frac{\pi}{N+1} \right) - \cos \left( \frac{N \pi}{N+1} \right) \biggl\},
\end{equation}
such that the normalized energy level spacings $t_m = S_m/D$ are completely defined via the formula
\begin{equation}
\label{eq:xdfdfsf}
   t_m = \left( N - 1 \right) \frac{ \cos \left( \pi \frac{N - m}{N+1} \right) - \cos \left( \pi \frac{N+1 - m}{N+1} \right)}{\cos \left( \frac{\pi}{N+1} \right) - \cos \left(  \frac{N \pi}{N+1} \right)}.
\end{equation}
There are $N-1$ values of the normalized energy level spacings $t_m$. For an odd number $N$ of oscillators in the chain, the values of $t_m$ for the reduced set $m = \{ 1, 2, ..., (N-1)/2 \}$ perfectly match with the values of $t_m$ obtained from the remaining set where $m = \{ N, N-1, ..., (N+1)/2 \}$. Therefore, since we are ultimately only interested in the thermodynamic limit of $N \to \infty$, we assume $N$ is odd and restrict the index of $t_m$ to $m = \{ 1, 2, ..., (N-1)/2 \}$ only in what follows. Since each normalized energy level spacing $t_m$ within this restricted index range of $m$ is both unique and runs in ascending order, the number of energy level spacings within some range $S$ increases from zero one-by-one after passing the increasing values of $t_1$, $t_2$, $t_3$ and so on with increasing $S$, up to the largest energy level spacing $t_{(N-1)/2}$. Hence the fraction of all energy level spacings (there are a total of $(N-1)/2$ energy level spacings in this analysis) with an energy level spacing below a certain $t_m$ is $C_m = 2(m-1)/(N-1)$. That is, there are zero energy level spacings with a spacing smaller than $t_1$ and so the fraction $C_1 = 0$, while there are $(N-3)/2$ energy level spacings with a spacing smaller than $t_{(N-1)/2}$ and so the fraction $C_{(N-1)/2} = 1 - 2/(N-1)$. Let us now take the thermodynamic limit of $N \to \infty$ with the normalized energy level spacings $t_m$ of Eq.~\eqref{eq:xdfdfsf}, which yields the simplification
\begin{equation}
\label{eq:xdsdfsdfsdfdfsf}
   t_m = \frac{\pi}{2} \sin \left( \frac{m \pi}{N+1} \right),
\end{equation}
which, due to the restricted index $m = \{ 1, 2, ..., (N-1)/2 \}$, has the bounds $0 \le t_m \le \pi/2$. Within this thermodynamic limit, the fractional counting quantity $C_m = 2(m-1)/(N-1)$ similarly satisfies the expected continuous range $0 \le C_m \le 1$. Then, upon explicitly eliminating the index $m$ in Eq.~\eqref{eq:xdsdfsdfsdfdfsf} with the help of $C_m$, and noting that we are working within the continuum limit of $N \to \infty$, we finally arrive at the desired continuum result
\begin{equation}
\label{eq:xdsdfsdfsdfdgrrgrgrgfsf}
   t = \frac{\pi}{2} \sin \left( \frac{\pi}{2} C \right),
\end{equation}
where we have replaced $t_m$ and $C_m$ with their continuous counterparts: the normalized energy level spacing $t$ and the cumulative probability $C$, which naturally fulfills the proper range $0 \le C \le 1$. The inversion of Eq.~\eqref{eq:xdsdfsdfsdfdgrrgrgrgfsf} provides the equation for the probability $C(t)$ directly, as was already quoted in Eq.~\eqref{eq:bomdaws}. We also sketch Eq.~\eqref{eq:bomdaws} with the cyan line in Fig.~\ref{wig}~(a), alongside the cumulative distribution functions of some other celebrated models. The picket fence model with equal energy level spacings necessitates a step function behaviour (red line), the uncorrelated energy levels model with random energy level spacings yields a smooth step function (green line), and a chaotic model described by a Gaussian orthogonal ensemble essentially exhibits a Gaussian function (yellow line)~\cite{Mehta2014, Guhr1998}.

The probability density function $p(s)$ associated with Eq.~\eqref{eq:xdsdfsdfsdfdgrrgrgrgfsf} was given in Eq.~\eqref{eq:bomdaws2}, and it is represented by the cyan line in Fig.~\ref{wig}~(b) along with the other distributions corresponding to the cumulative curves of Fig.~\ref{wig}~(a). Namely, the picket fence model has a Dirac delta function distribution (red line) due to the completely equal level spacings, while the uncorrelated model has a Poisson distribution (green line) implying that small level spacings are highly likely since $p(s \to 0) = 1$, while large level spacings are vanishingly unlikely with an exponential fall-off $p(s) = \mathrm{e}^{-s}$. The chaotic model has a Wigner distribution (yellow line) which suggests significant energy level repulsion due to the behaviour $p(s \to 0) = 0$, along with vanishing chances for large level spacings due to the very fast Gaussian decay $p(s \to \infty) \simeq \mathrm{e}^{- \pi s^2 /4}$. Interestingly, the inverse-square root distribution of the chain B model as described by Eq.~\eqref{eq:bomdaws2} (cyan line) demonstrates the absence of energy level repulsion since $p(s \to 0) = (2/\pi)^2 \simeq 0.405$, after which increasingly large level separations are increasingly likely up to the critical value of $s = \pi/2$. Above this value there is a brutal step function drop-off to zero due to the effective formation of a finite cosine band of energy levels [cf. Eq.~\eqref{eq:sfdfdfsdfsdsdbbbsdf}].
\\


\renewcommand{\theequation}{C \arabic{equation}}
\section{The chain A--B model}
\label{Sec:chainAB}

\begin{figure*}[tb]
 \includegraphics[width=\linewidth]{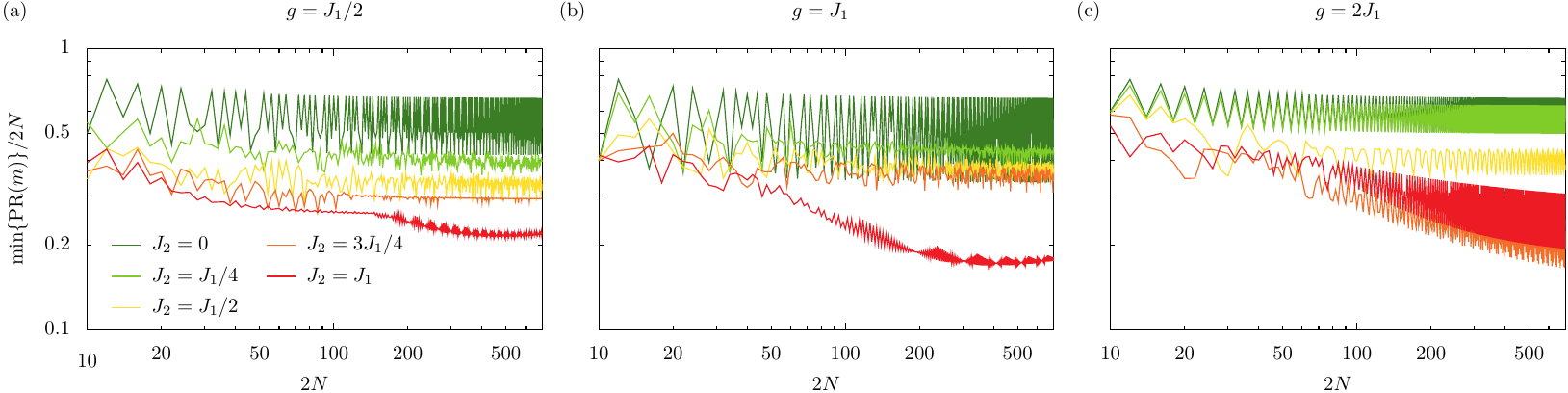}
 \caption{ \textbf{Finite size effects in the two-leg ladder model.} Log-log plots of the minimum (taken over all eigenstates $m$) of the scaled participation ratio $\mathrm{PR} \left( m \right) / 2 N$ as a function of $2N$, the total number of oscillators in the chain A--B system [cf. Eq.~\eqref{eq:sdfddwsdfs}]. We consider three values of the inter-chain coupling strength $g$ across the three panels of the figure, and five values of the next-nearest neighbour coupling strength $J_2$ in each case (as marked by the plot legend in the first panel). Panel (a): $g = J_1/2$. Panel (b): $g = J_1$. Panel (c): $g = 2 J_1$. This figure supplements the results of Fig.~\ref{numer}~(b) in effectively taking horizontal cuts through Fig.~\ref{numer}~(a) at three values of $g$, followed by vertical cuts at five values of $J_2$, and then extending the outcome as a function of the system size $2N$ (up to $2N = 700$).}
 \label{extraa}
\end{figure*}

In this appendix we house some additional calculations for a finite-sized two-leg ladder which, in the same manner as Fig.~\ref{numer}~(b), aids the interpretation of the overall participation ratio map for the chain A--B model as given in Fig.~\ref{numer}~(a). In essence, we take three horizontal cuts through the map of Fig.~\ref{numer}~(a) at certain values of the inter-chain coupling strength $g$, followed by vertical cuts at five values of the next-nearest neighbour coupling strength $J_2$, before we then extend the results as a function of the system size $2N$, the total number of oscillators in the system, to finally realize the desired plots. The outcomes are displayed in the three log-log plots of Fig.~\ref{extraa}, where we plot the minimum (taken over all eigenstates $m$) of the scaled participation ratio $\mathrm{min} \{ \mathrm{PR} \left( m \right) \} / 2 N$ as a function of $2N$ [cf. Eq.~\eqref{eq:sdfddwsdfs}].

We consider three increasingly large values of $g$ across the three panels of Fig.~\ref{extraa}, and we pick out five values of $J_2$ in each case (as marked by the plot legend in the first panel). Characteristically, across all of the panels in Fig.~\ref{extraa} we notice that when $J_2$ is sufficiently small (darker green, lighter green and yellow lines) all eigenstates $m$ have participation ratios $\mathrm{PR} \left( m \right)$ which scale essentially linearly with the system size $2N$, with only the prefactor changing from $\sim 0.7$ to $\sim 0.3$. In Fig.~\ref{extraa}~(a)~and~(b), where the inter-chain coupling strength $g = J_1/2$ and $g = J_1$ respectively, the cases of maximal next-nearest coupling $J_2 = J_1$ (red lines) display seemingly weakly sublinear scalings in this range of system size up to $2N = 700$, which is consistent with the interpretation of the participation ratio map of Fig.~\ref{numer}~(a). The protruding feature on the right-hand-side of Fig.~\ref{numer}~(a) is supported by the results of Fig.~\ref{extraa}~(c), where now $g = 2J_1$, which shows that the two largest values of $J_2$ (orange and red lines) are both associated with edge states rather than extended states, as implied by the apparent sublinear scalings of the participation ratio $\mathrm{PR} \left( m \right)$ with the system size $2N$ (for the values of $2N$ considered). We do not enter a discussion of even larger system sizes, since these cases fall outside of the current experimental scope of coupled systems with tunable next-nearest neighbour coupling~\cite{Periwal2021, ZhuGao2022, Martinez2021}. Finally, the highly non-monotonic data presented in Fig.~\ref{extraa} is strongly influenced by the exact value of the integer $2N$ describing the system size. We do not provide a commentary on the number theory of the presented model here, except to note the clear even-odd effect in the simplified chain B model [as discussed around Eq.~\eqref{eq:grgbrd}].
\\


\renewcommand{\theequation}{D \arabic{equation}}
\section{The chain A--B model with detuning}
\label{Sec:detune}

\begin{figure*}[tb]
 \includegraphics[width=\linewidth]{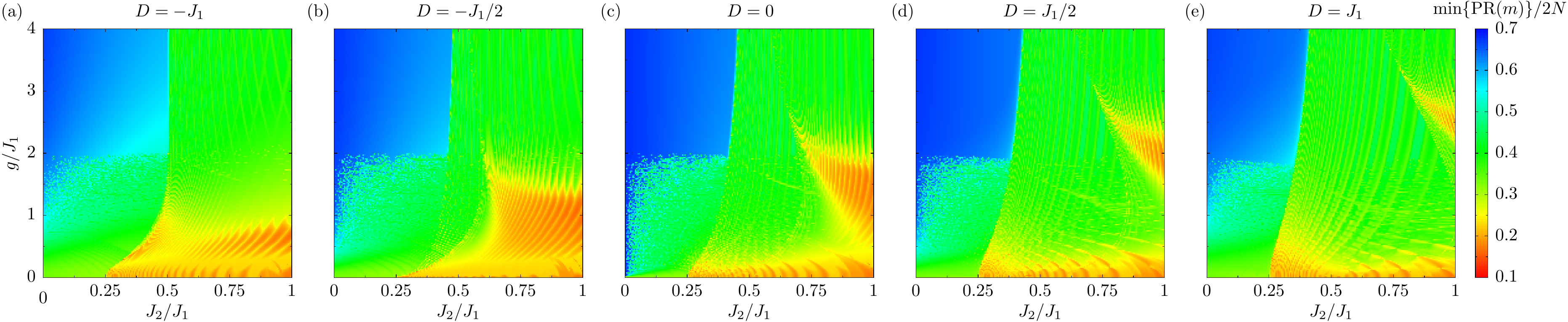}
 \caption{ \textbf{Degrees of localization in the two-leg ladder model with detuning.} Maps of the minimum of the scaled participation ratio $\mathrm{min} \{ \mathrm{PR} \left( m \right) \} / 2 N$, as a function of both the inter-chain coupling strength $g$ and the next-nearest neighbour coupling strength $J_2$, both in units of the nearest-neighbour coupling $J_1$ [cf. Eq.~\eqref{eq:sdfddwsdfs}]. We consider five different detunings $D = \omega_A - \omega_B$, defined via the resonance frequencies $\omega_A$ and $\omega_B$ of the oscillators in the upper chain and lower chain respectively, across the row of panels as marked as the top of the figure. The case of the central panel (c) recovers the model of the main text, where $D = 0$ (since $\omega_A = \omega_B = \omega_0$). In this figure, we consider a two-leg ladder with $2N = 200$ oscillators.}
 \label{NEGnumer}
\end{figure*}

In this appendix we consider a small generalization of the model of the main text in order to allow for a difference in resonance frequency between the oscillators of chain A and chain B. Instead of the common resonance frequency $\omega_0$ for all oscillators in the two-leg ladder [cf. Eq.~\eqref{eq:sdfsdfsfd} and Eq.~\eqref{eq:sdfsdssfdsfsfd}], we suppose that the resonators in the upper chain are associated with the frequency $\omega_A$, while those in the lower chain are instead describable with the frequency $\omega_B$. The inter-chain detuning $D$, where
\begin{equation}
\label{eq:dee}
  D = \omega_A - \omega_B,
\end{equation}
of the bipartite system then becomes a meaningful parameter. We provide some information about the degrees of localization in this slightly generalized model in Fig.~\ref{NEGnumer}, which presents participation ratio maps in the style of Fig.~\ref{numer}~(a) from the main text but now for different values of the detuning parameter $D$ (for the case of a two-leg ladder with $2N = 200$ oscillators). The central panel (c) in Fig.~\ref{NEGnumer} (where $D = 0$) repeats the zero detuning results of Fig.~\ref{numer}~(a), while panels (a) and (b) show the impact of negative detunings ($D < 0$). Panels (d) and (e) in Fig.~\ref{NEGnumer} similarly show the consequences of positive detunings ($D > 0$). Most noticeably, the principle features seen in Fig.~\ref{NEGnumer}~(c) are also observable in panels (a, b, d, e), but clearly it is preferable to have a negative detuning as in Fig.~\ref{NEGnumer}~(a, b) in order to create a larger red-orange zone where edge states reside.

\begin{figure*}[tb]
 \includegraphics[width=\linewidth]{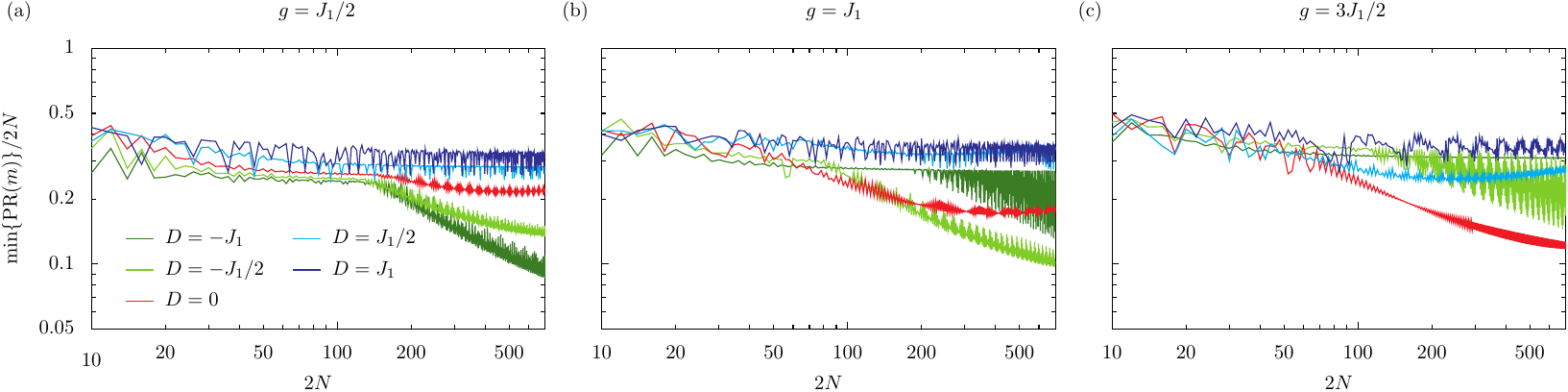}
 \caption{ \textbf{Finite size effects in the two-leg ladder model with detuning.} Log-log plots of the minimum (taken over all eigenstates $m$) of the scaled participation ratio $\mathrm{PR} \left( m \right) / 2 N$ as a function of $2N$, the total number of oscillators in the chain A--B system with detuning [cf. Eq.~\eqref{eq:sdfddwsdfs}]. We consider five values of the inter-chain detuning $D$ [as marked by the plot legend in panel (a)], corresponding to the values in  Fig.~\ref{NEGnumer}~(a--e). This figure supplements the results of Fig.~\ref{NEGnumer} by effectively taking horizontal cuts through Fig.~\ref{NEGnumer}~(a--e) at three values of $g$ [as marked at the top of this figure] and a single vertical cut at $J_2 = J_1$, and then extending the outcome as a function of the system size $2N$ (up to $2N = 700$).}
 \label{appy}
\end{figure*}

We track the finite-size effects associated with Fig.~\ref{NEGnumer} in Fig.~\ref{appy}, which takes cuts at $J_2 = J_1$ and at three values of the inter-chain coupling $g$ across Fig.~\ref{appy}~(a, b, c), for the five cases of detuning $D$ already considered in Fig.~\ref{NEGnumer}. The special case of zero detuning ($D = 0$) is shown in red, two negative detunings ($D < 0$) are represented with green lines and two positive detunings ($D > 0$) are displayed with blue lines. In Fig.~\ref{appy}~(a), where $g = J_1/2$, the cases with negative detuning (light and dark green lines) clearly display a sublinear scaling with the system size in this range up to $2N = 700$, as was already anticipatable from Fig.~\ref{NEGnumer}. When $g = J_1$ in Fig.~\ref{appy}~(b), the $D = -J_1/2$ case (light green line) is particularly sublinear, suggesting that some amount of detuning can be beneficial for the creation of edge states with a stronger degree of localization. Finally in Fig.~\ref{appy}~(c), where $g = 3J_1/2$, the zero detuning case (red line) shows gently oscillating sublinear behaviour, as opposed to the strong oscillations observable when $D = -J_1/2$ (light green line), which points to the non-trivial size-dependencies arising in the two-leg ladder model even for $2N \sim 10^3$. Importantly, the results presented in this appendix suggest that the particular flavour of edge state introduced in the main text is indeed robust against nonzero onsite energy detunings, improving the perspectives for future experimental detection.
\\
\\

\noindent \textbf{Acknowledgments}\\
\textit{Funding}: CAD is supported by the Royal Society via a University Research Fellowship (URF\slash R1\slash 201158) and an Exeter-FAPESP SPRINT grant with the Universidade Federal de São Carlos (state of São Paulo, Brazil). LMM acknowledges the support of Project PID2020-115221GB-C41, which was financed by MCIN/AEI/10.13039/501100011033 and the Aragón Government through Project Q-MAD. OIRF is funded by the EPSRC via the Maths DTP 2021-22 University of Exeter (EP/W523859/1) \textit{Discussions}: We thank R.~Bachelard, A. Cidrim, A. C. Santos, C. J. Villas-Boas and G.~Weick for fruitful discussions. We are indebted to Zhen Gao and Ce Shang for constructive discussions surrounding the experimental feasibility of our theoretical proposal, which comes in the wake of their innovative experimental work as described in Ref.~\cite{ZhuGao2022}. \textit{Hospitality}: CAD is grateful for the support of UFSCar during his visits to São Carlos. \textit{Data and materials availability}: All data is available inside the manuscript. The code is available at \href{https://github.com/of280/Ladder-Project}{https://github.com/of280/Ladder-Project} and is curated by O.~I.~R.~Fox. 
\\
\\


\noindent
\textbf{ORCID}\\
C.~A.~Downing: \href{https://orcid.org/0000-0002-0058-9746}{0000-0002-0058-9746}.
\\
L.~Martín-Moreno: \href{https://orcid.org/0000-0001-9273-8165}{0000-0001-9273-8165}.
\\




\begin{thebibliography}{100}



\bibitem{Anderson1958}
P.~W.~Anderson,
Absence of diffusion in certain random lattices,
\href{https://doi.org/10.1016/0370-1573(74)90029-5}
{Phys. Rev. \textbf{109}, 1492 (1958)}.

\bibitem{Thouless1974}
D.~J.~Thouless,
Electrons in disordered systems and the theory of localization,
\href{https://doi.org/10.1016/0370-1573(74)90029-5}
{Phys. Rep. \textbf{13}, 93 (1974)}.


\bibitem{Wiersma2013}
D.~S.~Wiersma,
Disordered photonics,
\href{https://doi.org/10.1038/nphoton.2013.29}
{Nat. Photonics \textbf{7}, 188 (2013)}.

\bibitem{Segev2013}
M.~Segev, Y.~Silberberg and D.~N.~Christodoulides,
Anderson localization of light,
\href{https://doi.org/10.1038/nphoton.2013.30}
{Nat. Photonics \textbf{7}, 197 (2013)}.


\bibitem{Wannier1962}
G.~H.~Wannier,
Dynamics of band electrons in electric and magnetic fields,
\href{https://doi.org/10.1103/RevModPhys.34.645}
{Rev. Mod. Phys. \textbf{34}, 645 (1962)}.

\bibitem{Emin1987}
D.~Emin and C.~F.~Hart,
Existence of Wannier-Stark localization,
\href{https://doi.org/10.1103/PhysRevB.36.7353}
{Phys. Rev. B \textbf{36}, 7353 (1987)}.


\bibitem{Arikawa2009}
M.~Arikawa, S.~Tanaya, I.~Maruyama and Y.~Hatsugai,
Edge states of a spin-$1/2$ two-leg ladder with four-spin ring exchange,
\href{https://doi.org/10.1103/PhysRevB.79.205107}
{Phys. Rev. B \textbf{79}, 205107 (2009)}.

\bibitem{Hasan2010}
M.~Z.~Hasan and C.~L.~Kane,
Colloquium: topological insulators,
\href{https://doi.org/10.1103/RevModPhys.82.3045}
{Rev. Mod. Phys. \textbf{82}, 3045 (2010)}.


\bibitem{Asboth2016}
J.~K.~Asboth, L.~Oroszlany and A.~Palyi,
\textit{A Short Course on Topological Insulators},
(Springer, London, 2016)
and
\href{https://arxiv.org/abs/1509.02295}
{arXiv:1509.02295}.


\bibitem{Hsu2016}
C.~W.~Hsu, B.~Zhen, A.~D.~Stone, J.~D.~Joannopoulos and M.~Soljačić,
Bound states in the continuum,
\href{https://doi.org/10.1038/natrevmats.2016.48}
{Nat. Rev. Mater. \textbf{1}, 16048 (2016)}.


\bibitem{Azzam2020}
S.~I.~Azzam and A.~V.~Kildishev,
Photonic bound states in the continuum: from basics to applications,
\href{https://doi.org/10.1002/adom.202001469}
{Rep. Prog. Phys. \textbf{9}, 2001469 (2020)}.


\bibitem{Sadreev2021}
A.~F~Sadreev,
Interference traps waves in an open system: bound states in the continuum,
\href{https://doi.org/10.1088/1361-6633/abefb9}
{Rep. Prog. Phys. \textbf{84}, 055901 (2021)}.

\bibitem{Koshelev2023}
K.~L.~Koshelev, Z.~F.~Sadrieva, A.~A.~Shcherbakov, Y.~S.~Kivshar and A.~A.~Bogdanov,
Bound states in the continuum in photonic structures,
\href{https://doi.org/10.3367/UFNe.2021.12.039120}
{Phys.-Uspekhi \textbf{66}, 494 (2023)}.


\bibitem{Neumann1929}
J.~von~Neumann and E.~P.~Wigner,
Über merkwürdige diskrete eigenwerte,
\href{https://doi.org/10.3367/UFNe.2021.12.039120}
{Phys. Z. \textbf{30}, 465 (1929)}.


\bibitem{Fairbairn1968}
W.~M.~Fairbairn,
Surface states in the linear chain with next-nearest-neighbour interactions,
\href{https://doi.org/10.1016/0039-6028(68)90147-7}
{Surf. Sci. \textbf{9}, 439 (1968)}.



\bibitem{Fisher1981}
M.~E.~Fisher and W.~Selke,
Low temperature analysis of the axial next-nearest neighbour Ising model near its multiphase point,
\href{https://doi.org/10.1098/rsta.1981.0156}
{Phil. Trans. R. Soc. A \textbf{302}, 1 (1981)}.



\bibitem{Landau1983}
D.~P.~Landau,
Critical and multicritical behavior in a triangular-lattice-gas Ising model: repulsive nearest-neighbor and attractive next-nearest-neighbor coupling,
\href{https://doi.org/10.1103/PhysRevB.27.5604}
{Phys. Rev. B \textbf{27}, 5604 (1983)}.

\bibitem{Han1994}
J.~H.~Han, D.~J.~Thouless, H.~Hiramoto and M.~Kohmoto,
Critical and bicritical properties of Harper’s equation with next-nearest-neighbor coupling,
\href{https://doi.org/10.1103/PhysRevB.50.11365}
{Phys. Rev. B \textbf{50}, 11365 (1994)}.


\bibitem{Chandler2016}
C.~J.~Chandler, C.~Prosko and F.~Marsiglio,
The effect of next-nearest neighbour hopping in the one, two, and three dimensional Holstein model,
\href{https://doi.org/10.1038/srep32591 }
{Sci. Rep. \textbf{6}, 32591 (2016)}.


\bibitem{Laha2017}
A.~Laha, A.~Biswas and S.~Ghosh,
Next-nearest-neighbor resonance coupling and exceptional singularities in degenerate optical microcavities,
\href{https://doi.org/10.1364/JOSAB.34.002050}
{J. Opt. Soc. Am. B \textbf{34}, 2050 (2017)}.


\bibitem{Mittal2018}
D.~Leykam, S.~Mittal, M.~Hafezi and Y.~D.~Chong,
Reconfigurable topological phases in next-nearest-neighbor coupled resonator lattices,
\href{https://doi.org/10.1103/PhysRevLett.121.023901}
{Phys. Rev. Lett. \textbf{121}, 023901 (2018)}.

\bibitem{Chen2021}
Y.~Chen, M.~Kadic and M.~Wegener,
Roton-like acoustical dispersion relations in 3D metamaterials,
\href{https://doi.org/10.1038/s41467-021-23574-2}
{Nat. Commun. \textbf{12}, 3278 (2021)}.

\bibitem{Saroka2021}
C.~A.~Downing and V.~A.~Saroka,
Exceptional points in oligomer chains,
\href{https://doi.org/10.1038/s42005-021-00757-3}
{Commun. Phys. \textbf{4}, 254 (2021)}.

\bibitem{Aristov2022}
D.~Aristov, H.~Sigurdsson and P.~G.~Lagoudakis,
Screening nearest-neighbor interactions in networks of exciton-polariton condensates through spin-orbit coupling,
\href{https://doi.org/10.1103/PhysRevB.105.155306}
{Phys. Rev. B \textbf{105}, 155306 (2022)}.

\bibitem{Wang2022}
K.~Wang, Y.~Chen, M.~Kadic, C.~Wang and M.~Wegener,
Nonlocal interaction engineering of 2D roton-like dispersion relations in acoustic and mechanical metamaterials,
\href{https://doi.org/10.1038/s43246-022-00257-z}
{Commun. Mater. \textbf{3}, 35 (2022)}.

\bibitem{Kazemi2023}
A.~Kazemi, K.~J.~Deshmukh, F.~Chen, Y.~Liu, B.~Deng, H.~C.~Fu and P.~Wang,
Drawing dispersion curves: band structure customization via nonlocal phononic crystals,
\href{https://doi.org/10.1103/PhysRevLett.131.176101}
{Phys. Rev. Lett. \textbf{131}, 176101 (2023)}.



\bibitem{Periwal2021}
A.~Periwal, E.~S.~Cooper, P.~Kunkel, J.~F.~Wienand, E.~J.~Davis and M.~Schleier-Smith,
Programmable interactions and emergent geometry in an array of atom clouds,
\href{https://doi.org/10.1038/s41586-022-04610-7}
{Nature \textbf{600}, 630 (2021)}.


\bibitem{ZhuGao2022}
Z.~Zhu, Z.~Gao, G.-G.~Liu, Y.~Ge, Y.~Wang, X.~Xi, B.~Yan, F.~Chen, P.~P.~Shum and H.-X.~Sun,
Observation of multiple rotons and multidirectional roton-like dispersion relations in acoustic metamaterials,
\href{https://doi.org/10.1088/1367-2630/aca786}
{New J. Phys. \textbf{24}, 123019 (2022)}.


\bibitem{Martinez2021}
J.~A.~Iglesias~Martínez, M.~F.~Groß, Y.~Chen, T.~Frenzel, V.~Laude, M.~Kadic and M.~Wegener,
Experimental observation of roton-like dispersion relations in metamaterials,
\href{https://doi.org/10.1126/sciadv.abm2189}
{Sci. Adv. \textbf{7}, eabm2189 (2021)}.

\bibitem{Bossart2023}
A.~Bossart and R.~Fleury,
Extreme spatial dispersion in nonlocally resonant elastic metamaterials,
\href{https://doi.org/10.1103/PhysRevLett.130.207201}
{Phys. Rev. Lett. \textbf{130}, 207201 (2023)}.

\bibitem{Rajabpoor2023}
A.~R.~Alisepahi, S.~Sarkar, K.~Sun and J.~Ma,
Breakdown of conventional winding number calculation in one-dimensional lattices with interactions beyond nearest neighbors,
\href{https://arxiv.org/abs/2304.04080}
{arXiv:2304.04080}.


\bibitem{Caselli2015}
N.~Caselli, F.~Riboli, F.~La~China, A.~Gerardino, L.~Li, E.~H.~Linfield, F.~Pagliano, A.~Fiore, F.~Intonti and M.~Gurioli,
Tailoring the photon hopping by nearest-neighbor and next-nearest-neighbor interaction in photonic arrays,
\href{https://doi.org/10.1021/acsphotonics.5b00041}
{ACS Photonics \textbf{2}, 565 (2015)}.

\bibitem{Senanian2023}
A.~Senanian, L.~G.~Wright, P.~F.~Wade, H.~K.~Doyle and P.~L.~McMahon,
Programmable large-scale simulation of bosonic transport in optical synthetic frequency lattices,
\href{https://doi.org/10.1038/s41567-023-02075-7}
{Nat. Phys. \textbf{19}, 1333 (2023)}.



\bibitem{Alyatkin2020}
S. Alyatkin, J.~D. Topfer, A.~Askitopoulos, H.~Sigurdsson and P.~G.~Lagoudakis,
Optical control of couplings in polariton condensate lattices,
\href{https://doi.org/10.1103/PhysRevLett.124.207402}
{Phys. Rev. Lett. \textbf{124}, 207402 (2020)}.


\bibitem{Dovzhenko2023}
D.~Dovzhenko, D.~Aristov, L.~Pickup, H.~Sigurdsson and P.~G.~Lagoudakis,
Next nearest neighbour coupling with spinor polariton condensates,
\href{https://arxiv.org/abs/2301.04210}
{arXiv:2301.04210}.


\bibitem{Troyer1996}
M.~Troyer, H.~Tsunetsugu and T.~M.~Rice,
Properties of lightly doped $t-J$ two-leg ladders,
\href{https://doi.org/10.1103/PhysRevB.53.251}
{Phys. Rev. B \textbf{53}, 251 (1996)}.


\bibitem{Donohue2001}
P.~Donohue and T.~Giamarchi,
Mott--superfluid transition in bosonic ladders,
\href{https://doi.org/10.1103/PhysRevB.63.180508}
{Phys. Rev. B \textbf{63}, 180508(R) (2001)}.

\bibitem{Orignac2001}
E.~Orignac and T.~Giamarchi,
Meissner effect in a bosonic ladder,
\href{https://doi.org/10.1103/PhysRevB.64.144515}
{Phys. Rev. B \textbf{64}, 144515 (2001)}.


\bibitem{Robinson2012}
N.~J.~Robinson, F.~H.~L.~Essler, E.~Jeckelmann and A.~M.~Tsvelik,
Finite wave vector pairing in doped two-leg ladders,
\href{https://doi.org/10.1103/PhysRevB.85.195103}
{Phys. Rev. B \textbf{85}, 195103 (2012)}.


\bibitem{Wei2014}
R.~Wei and E.~J.~Mueller,
Theory of bosons in two-leg ladders with large magnetic fields,
\href{https://doi.org/10.1103/PhysRevA.89.063617}
{Phys. Rev. A \textbf{89}, 063617 (2014)}.


\bibitem{Ye2022}
R.~Ye, G.~Li, L.~Wang, X.~Wu, L.~Yuan and X.~Chen,
Controlling localized states in a two-leg ladder lattice with diagonal edges via gain/loss,
\href{https://doi.org/10.1364/OME.477926}
{Opt. Mater. Express \textbf{12}, 4755 (2022)}.


\bibitem{Atala2014}
M.~Atala, M.~Aidelsburger, M.~Lohse, J.~T.~Barreiro, B.~Paredes and I.~Bloch,
Observation of chiral currents with ultracold atoms in bosonic ladders,
\href{https://doi.org/10.1038/nphys2998}
{Nature Phys. \textbf{10}, 588 (2014)}.

\bibitem{Greschner2015}
S.~Greschner, M.~Piraud, F.~Heidrich-Meisner, I.~P.~McCulloch, U.~Schollwöck and T.~Vekua,
Spontaneous increase of magnetic flux and chiral--current reversal in bosonic ladders: swimming against the tide,
\href{https://doi.org/10.1103/PhysRevLett.115.190402}
{Phys. Rev. Lett. \textbf{115}, 190402 (2015)}.

\bibitem{Tai2017}
M.~E.~Tai, A.~Lukin, M.~Rispoli, R.~Schittko, T.~Menke, D.~Borgnia, P.~M.~Preiss, F.~Grusdt, A.~M.~Kaufman and M.~Greiner,
Microscopy of the interacting Harper–Hofstadter model in the two-body limit,
\href{https://doi.org/10.1038/nature22811}
{Nature \textbf{546}, 519 (2017)}.

\bibitem{An2017}
F.~A.~An, E.~J.~Meier and B.~Gadway,
Direct observation of chiral currents and magnetic reflection in atomic flux lattices,
\href{https://doi.org/10.1126/sciadv.1602685}
{Sci. Adv. \textbf{3}, e1602685 (2017)}.


\bibitem{Zhu2022}
Q.~Zhu, Z.-H.~Sun, M.~Gong, F.~Chen, Y.-R.~Zhang, Y.~Wu, Y.~Ye, C.~Zha, S.~Li, S.~Guo, H.~Qian, H.-L.~Huang, J.~Yu, H.~Deng, H.~Rong, J.~Lin, Y.~Xu, L.~ Sun, C.~Guo, N.~Li, F.~Liang, C.-Z.~Peng, H.~Fan, X.~Zhu and J.-W.~Pan,
Observation of thermalization and information scrambling in a superconducting quantum processor,
\href{https://doi.org/10.1103/PhysRevLett.128.160502}
{Phys. Rev. Lett. \textbf{128}, 160502 (2022)}.

\bibitem{Xiao2023}
Y.~Li, H.~Du, Y.~Wang, J.~Liang, L.~Xiao, W.~Yi, J.~Ma and S.~Jia,
Observation of frustrated chiral dynamics in an interacting triangular flux ladder,
\href{https://doi.org/10.1038/s41467-023-43204-3}
{Nat. Commun. \textbf{14}, 7560 (2023)}.



\bibitem{Deng2015}
H.~Deng, H.~Dai, J.~Huang, X.~Qin, J.~Xu, H.~Zhong, C.~He and C.~Lee,
Cluster Gutzwiller study of the Bose-Hubbard ladder: ground-state phase diagram and many-body Landau-Zener dynamics,
\href{https://doi.org/10.1103/PhysRevA.92.023618}
{Phys. Rev. A \textbf{92}, 023618 (2015)}.


\bibitem{Qiao2021}
X.~Qiao, X.-B.~Zhang, Y.~Jian, A.-X.~Zhang, Z.-F.~Yu and J.-K.~Xue,
Quantum phases of interacting bosons on biased two-leg ladders with magnetic flux,
\href{https://doi.org/10.1103/PhysRevA.104.053323}
{Phys. Rev. A \textbf{104}, 053323 (2021)}.


\bibitem{Padhan2023}
A.~Padhan, R.~Parida, S.~Lahiri, M.~K.~Giri and T.~Mishra,
Quantum phases of constrained bosons on a two-leg Bose-Hubbard ladder,
\href{https://doi.org/10.1103/PhysRevA.108.013316}
{Phys. Rev. A \textbf{108}, 013316 (2023)}.

\bibitem{Fan2023}
J.~Fan, X.~Zhou and S.~Jia,
Quantum phases of the biased two-chain-coupled Bose-Hubbard ladder,
\href{https://arxiv.org/abs/2308.15042}
{arXiv:2308.15042}.

\bibitem{Brillouin1953}
L.~Brillouin,
\textit{Wave Propagation in Periodic Structures: Electric Filters and Crystal Lattices},
(Dover, New York, 1953).



\bibitem{Altland1997}
A.~Altland and M.~R.~Zirnbauer,
Nonstandard symmetry classes in mesoscopic normal-superconducting hybrid structures,
\href{https://doi.org/10.1103/PhysRevB.55.1142}
{Phys. Rev. B \textbf{55}, 1142 (1997)}.


\bibitem{Velasco2019}
C.~G.~Velasco and B.~Paredes,
Classification of topological ladder models,
\href{https://arxiv.org/abs/1907.11460}
{arXiv:1907.11460}.


\bibitem{Bell1970}
R.~J.~Bell and P.~Dean,
Atomic vibrations in vitreous silica,
\href{https://doi.org/10.1039/DF9705000055}
{Discuss. Faraday Soc. \textbf{50}, 55 (1970)}.

\bibitem{Thouless1972}
J.~T.~Edwards and D.~J.~Thouless,
Electrons in disordered systems and the theory of localization,
\href{https://doi.org/10.1088/0022-3719/5/8/007}
{J. Phys. C: Solid State Phys. \textbf{5}, 807 (1972)}.

\bibitem{Torres2019}
E.~J.~Torres-Herrera, J.~A.~Méndez-Bermúdez and L.~F.~Santos,
Level repulsion and dynamics in the finite one-dimensional Anderson model,
\href{https://doi.org/10.1103/PhysRevE.100.022142}
{Phys. Rev. E \textbf{100}, 022142 (2019)}.

\bibitem{Schiulaz2018}
M.~Schiulaz, M.~Távora and L.~F.~Santos,
From few- to many-body quantum systems,
\href{https://doi.org/10.1088/2058-9565/aad913}
{Quantum Sci. Technol. \textbf{3}, 044006 (2018)}.



\bibitem{Sarkar2021}
M.~Sarkar, R.~Ghosh, A.~Sen and K.~Sengupta,
Mobility edge and multifractality in a periodically driven Aubry-André model,
\href{https://doi.org/10.1103/PhysRevB.103.184309}
{Phys. Rev. B \textbf{103}, 184309 (2021)}.


\bibitem{Duthie2022}
A.~Duthie, S.~Roy, and D.~E.~Logan,
Anomalous multifractality in quantum chains with strongly correlated disorder,
\href{https://doi.org/10.1103/PhysRevB.106.L020201}
{Phys. Rev. B \textbf{106}, L020201 (2022)}.


\bibitem{Aditya2023}
S.~Aditya, K.~Sengupta and D.~Sen,
Periodically driven model with quasiperiodic potential and staggered hopping amplitudes: engineering of mobility gaps and multifractal states,
\href{https://doi.org/10.1103/PhysRevB.107.035402}
{Phys. Rev. B \textbf{107}, 035402 (2023)}.


\bibitem{Mehta2014}
M.~L.~Mehta,
\textit{Random Matrices and the Statistical Theory of Energy Levels},
(Academic Press, London, 1967).


\bibitem{Guhr1998}
T.~Guhr, A.~Müller–Groeling and H.~A.~Weidenmüller,
Random-matrix theories in quantum physics: common concepts,
\href{https://doi.org/10.1016/S0370-1573(97)00088-4}
{Phys. Rep. \textbf{299}, 189 (1998)}.


\bibitem{Oganesyan2007}
V.~Oganesyan and D.~A.~Huse,
Localization of interacting fermions at high temperature,
\href{https://doi.org/10.1103/PhysRevB.75.155111}
{Phys. Rev. B \textbf{75}, 155111 (2007)}.

\bibitem{Atas2013}
Y.~Y.~Atas, E.~Bogomolny, O.~Giraud and G.~Roux,
Distribution of the ratio of consecutive level spacings in random matrix ensembles,
\href{https://doi.org/10.1103/PhysRevLett.110.084101}
{Phys. Rev. Lett. \textbf{110}, 084101 (2013)}.


\bibitem{Cuevas2012}
E.~Cuevas, M.~Feigel'man, L.~Ioffe and M.~Mezard,
Level statistics of disordered spin-1/2 systems and materials with localized Cooper pairs,
\href{https://doi.org/10.1038/ncomms2115}
{Nat. Commun. \textbf{3}, 1128 (2012)}.


\bibitem{Agarwal2015}
K.~Agarwal, S.~Gopalakrishnan, M.~Knap, M.~Müller and E.~Demler,
Anomalous diffusion and Griffiths effects near the many-body localization transition,
\href{https://doi.org/10.1103/PhysRevLett.114.160401}
{Phys. Rev. Lett. \textbf{114}, 160401 (2015)}.


\bibitem{Morong2021}
W.~Morong, F.~Liu, P.~Becker, K.~S.~Collins, L.~Feng, A.~Kyprianidis, G.~Pagano, T.~You, A.~V.~Gorshkov and C.~Monroe,
Observation of Stark many-body localization without disorder,
\href{https://doi.org/10.1038/s41586-021-04271-y}
{Nature \textbf{599}, 393 (2021)}.


\bibitem{Downing2021}
C.~A.~Downing and L.~Martín-Moreno,
Polaritonic Tamm states induced by cavity photons,
\href{https://doi.org/10.1515/nanoph-2020-0370}
{Nanophotonics \textbf{10}, 513 (2021)}.


\bibitem{Zhou2022}
Z.~Zhou, S.~Huang, D.~Li, J.~Zhu and Y.~Li,
Broadband impedance modulation via non-local acoustic metamaterials,
\href{https://doi.org/10.1093/nsr/nwab171}
{Natl. Sci .Rev. \textbf{9}, nwab171 (2022)}.


\bibitem{Allard2022}
T.~F.~Allard and G.~Weick,
Disorder-enhanced transport in a chain of lossy dipoles strongly coupled to cavity photons,
\href{https://doi.org/10.1103/PhysRevB.106.245424}
{Phys. Rev. B \textbf{106}, 245424 (2022)}.

\bibitem{Allard2023}
T.~F.~Allard and G.~Weick,
Multiple polaritonic edge states in a Su-Schrieffer-Heeger chain strongly coupled to a multimode cavity,
\href{https://doi.org/10.1103/PhysRevB.108.245417}
{Phys. Rev. B \textbf{108}, 245417 (2023)}.


\bibitem{John1991}
S.~John,
Localization of light,
\href{https://doi.org/10.1063/1.881300}
{Phys.Today \textbf{44}, 32 (1991)}.


\bibitem{Lamata2019}
P.~Forn-Diaz, L.~Lamata, E.~Rico, J.~Kono and E.~Solano,
Ultrastrong coupling regimes of light-matter interaction,
\href{https://doi.org/10.1103/RevModPhys.91.025005}
{Rev. Mod. Phys. \textbf{91}, 025005 (2019)}.


\bibitem{Toghill2022}
C.~A.~Downing and A.~J.~Toghill,
Quantum topology in the ultrastrong coupling regime,
\href{https://doi.org/10.1038/s41598-022-15735-0}
{Sci. Rep. \textbf{12}, 11630 (2022)}.


\bibitem{Zurita2019}
J.~Zurita, C.~E.~Creffield and G.~Platero,
Topology and interactions in the photonic Creutz and Creutz-Hubbard ladders,
\href{https://doi.org/10.1002/qute.201900105}
{Adv. Quantum Technol. \textbf{3}, 1900105 (2021)}.

\bibitem{Azcona2021}
P.~Martínez~Azcona and C.~A.~Downing,
Doublons, topology and interactions in a one-dimensional lattice,
\href{https://doi.org/10.1038/s41598-021-91778-z}
{Sci. Rep. \textbf{11}, 12540 (2021)}.


\bibitem{Zimmermann1987}
T.~Zimmermann, H.~Köppel, E.~Haller, H.-D.~Meyer and L.~S.~Cederbaum,
Energy level statistics of coupled oscillators,
\href{https://doi.org/10.1088/0031-8949/35/2/006}
{Phys. Scr. \textbf{35}, 125 (1987)}.

\bibitem{Pandey1991}
A.~Pandey and R.~Ramaswamy,
Level spacings for harmonic-oscillator systems,
\href{https://doi.org/10.1103/PhysRevA.43.4237}
{Phys. Rev. A \textbf{43}, 4237 (1991)}.


\bibitem{Chakrabarti2003}
B.~Chakrabarti and B.~Hu,
Level correlation in coupled harmonic oscillator systems,
\href{https://doi.org/10.1016/S0375-9601(03)01001-6}
{Phys. Lett. A \textbf{315}, 93 (2003)}.


\bibitem{Tarquini2017}
E.~Tarquini, G.~Biroli and M.~Tarzia,
Critical properties of the Anderson localization transition and the high-dimensional limit,
\href{https://doi.org/10.1103/PhysRevB.95.094204}
{Phys. Rev. B \textbf{95}, 094204 (2017)}.






\end{thebibliography}
\end{document}